\documentclass[twocolumn]{aastex631}

\shorttitle{GHOST Emission Spectroscopy}
\shortauthors{Deibert et al.}
\graphicspath{{./}{figures/}}

\usepackage{printlen}
\usepackage{rotating}

\begin{document}

\title{High-Resolution Dayside Spectroscopy of WASP-189b: \\ Detection of Iron during the GHOST/Gemini South System Verification Run}

\author[0000-0001-9796-2158]{Emily K.~Deibert}
\affiliation{Gemini Observatory/NSF’s NOIRLab, Casilla 603, La Serena, Chile}
\correspondingauthor{Emily K.~Deibert}
\email{emily.deibert@noirlab.edu}

\author[0000-0002-4451-1705]{Adam B.~Langeveld}
\affil{Department of Astronomy and Carl Sagan Institute, Cornell University, Ithaca, New York 14853, USA}

\author[0000-0003-0672-7123]{Mitchell E.~Young}
\affil{Astrophysics Research Centre, Queen's University Belfast, Belfast BT7 1NN, UK}

\author[0000-0001-6362-0571]{Laura Flagg}
\affil{Department of Astronomy and Carl Sagan Institute, Cornell University, Ithaca, New York 14853, USA}

\author[0000-0001-7836-1787]{Jake D.~Turner}
\affil{Department of Astronomy and Carl Sagan Institute, Cornell University, Ithaca, New York 14853, USA}
\affil{NHFP Sagan Fellow}

\author[0000-0002-9946-5259]{Peter C.~B.~Smith}
\affil{School of Earth and Space Exploration, Arizona State University, Tempe, AZ, USA}

\author[0000-0001-6391-9266]{Ernst J.~W.~de Mooij}
\affil{Astrophysics Research Centre, Queen's University Belfast, Belfast BT7 1NN, UK}

\author[0000-0001-5349-6853]{Ray Jayawardhana}   
\affil{Department of Physics and Astronomy, Johns Hopkins University, 3400 N Charles Street, Baltimore, MD 21218, USA}

\author[0000-0002-9020-5004]{Kristin Chiboucas}   
\affiliation{Gemini Observatory/NSF’s NOIRLab, 670 N. A’ohoku Place, Hilo, HI, 96720, USA}

\author[0000-0002-5227-9627]{Roberto Gamen}  
\affiliation{Instituto de Astrof\'isica de La Plata, CONICET--UNLP, Paseo del Bosque s/n, 1900, La Plata, Argentina}

\author[0000-0003-2969-2445]{Christian R.~Hayes}
\affiliation{Space Telescope Science Institute, 3700 San Martin Drive, Baltimore, MD 21218, USA}

\author[0000-0003-2530-3000]{Jeong-Eun Heo}  
\affiliation{Gemini Observatory/NSF’s NOIRLab, Casilla 603, La Serena, Chile}

\author{Miji Jeong}  
\affiliation{Department of Astronomy, Space Science, and Geology, Chungnam National University, Daejeon 34134, Republic of Korea}

\author{Venu Kalari}   
\affiliation{Gemini Observatory/NSF’s NOIRLab, Casilla 603, La Serena, Chile}

\author[0000-0002-5084-168X]{Eder Martioli}   
\affiliation{Laborat\'orio Nacional de Astrof\'isica, Rua Estados Unidos 154, 37504-364, Itajub\'a, MG, Brazil}

\author[0000-0003-4479-1265]{Vinicius M.\ Placco}   
\affiliation{NSF’s NOIRLab, Tucson, AZ 85719, USA}

\author{Siyi Xu}   
\affiliation{Gemini Observatory/NSF’s NOIRLab, 670 N. A’ohoku Place, Hilo, HI, 96720, USA}

\author{Ruben Diaz}   
\affiliation{Gemini Observatory/NSF’s NOIRLab, Casilla 603, La Serena, Chile}

\author{Manuel Gomez-Jimenez}   
\affiliation{Gemini Observatory/NSF’s NOIRLab, Casilla 603, La Serena, Chile}

\author[0000-0001-5558-6297]{Carlos Quiroz}   
\affiliation{Gemini Observatory/NSF’s NOIRLab, Casilla 603, La Serena, Chile}

\author[0000-0001-7518-1393]{Roque Ruiz-Carmona}   
\affiliation{Gemini Observatory/NSF’s NOIRLab, Casilla 603, La Serena, Chile}

\author[0000-0001-8589-4055]{Chris Simpson}   
\affiliation{Gemini Observatory/NSF’s NOIRLab, 670 N. A’ohoku Place, Hilo, HI, 96720, USA}

\author[0000-0003-4666-6564]{Alan W.~McConnachie}  
\affiliation{NRC Herzberg Astronomy and Astrophysics Research Centre, 5071 West Saanich Road, Victoria, B.C., V9E 2E7, Canada}

\author{John Pazder}   
\affiliation{NRC Herzberg Astronomy and Astrophysics Research Centre, 5071 West Saanich Road, Victoria, B.C., V9E 2E7, Canada}
\affiliation{Department of Physics and Astronomy, University of Victoria, Victoria, BC, V8W 3P2, Canada}

\author[0000-0001-7548-5354]{Gregory Burley}   
\affiliation{NRC Herzberg Astronomy and Astrophysics Research Centre, 5071 West Saanich Road, Victoria, B.C., V9E 2E7, Canada}

\author[0000-0002-6194-043X]{Michael Ireland}  
\affiliation{Research School of Astronomy and Astrophysics, Australian National University, Canberra 2611, Australia}

\author{Fletcher Waller}   
\affiliation{Department of Physics and Astronomy, University of Victoria, Victoria, BC, V8W 3P2, Canada}

\author[0000-0002-2606-5078]{Trystyn A.~M.~Berg}   
\affiliation{NRC Herzberg Astronomy and Astrophysics Research Centre, 5071 West Saanich Road, Victoria, B.C., V9E 2E7, Canada}

\author[0000-0001-5528-7801]{J.~Gordon Robertson}   
\affiliation{Australian Astronomical Optics, Macquarie University, 105 Delhi Rd, North Ryde NSW 2113, Australia}
\affiliation{Sydney Institute for Astronomy, School of Physics, University of Sydney, NSW 2006, Australia}

\author[0000-0002-6230-0151]{David O.~Jones}   
\affiliation{Gemini Observatory/NSF’s NOIRLab, 670 N. A’ohoku Place, Hilo, HI, 96720, USA}
\affiliation{Institute for Astronomy, University of Hawai‘i, 640 N.\ Aohoku Pl., Hilo, HI 96720, USA}

\author[0000-0002-6633-7891]{Kathleen Labrie}   
\affiliation{Gemini Observatory/NSF’s NOIRLab, 670 N. A’ohoku Place, Hilo, HI, 96720, USA}

\author{Susan Ridgway}   
\affiliation{NSF’s NOIRLab, Tucson, AZ 85719, USA}

\author[0000-0003-1033-4402]{Joanna Thomas-Osip}   
\affiliation{Gemini Observatory/NSF’s NOIRLab, Casilla 603, La Serena, Chile}

\begin{abstract}
With high equilibrium temperatures and tidally locked {rotation}, ultra-hot Jupiters (UHJs) are unique laboratories within which to probe extreme atmospheric physics and chemistry. In this paper, we present high-resolution dayside spectroscopy of the UHJ WASP-189b obtained with the new Gemini High-resolution Optical SpecTrograph (GHOST) at the Gemini South Observatory. The observations, which cover three hours of post-eclipse orbital phases, were obtained during the instrument's System Verification run.
We detect the planet's atmosphere via the Doppler cross-correlation technique, and recover a detection of neutral iron in the planet's dayside atmosphere at a significance of {7.5}$\sigma$ in the red-arm of the data, verifying the presence of a thermal inversion.
We also investigate the presence of other species in the atmosphere and discuss the implications of model injection/recovery tests. These results represent the first atmospheric characterization of an exoplanet with GHOST's high-resolution mode, and demonstrate the potential of this new instrument in detecting and studying ultra-hot exoplanet atmospheres.
\end{abstract}

\section{Introduction}
\label{sec:intro}

Due to their high dayside temperatures, short orbital periods, and strong stellar irradiation levels, tidally locked ultra-hot Jupiters (UHJS; T${}_\mathrm{eq}\gtrsim$ 2200~K; \citealt{Parmentier18}) are unique laboratories within which to study a range of physical, chemical, and dynamical processes in giant exoplanet atmospheres. Under these extreme conditions, their highly irradiated permanent daysides can reach temperatures upwards of 3000~K \citep{Parmentier18} and exhibit different atmospheric chemistry from their cooler, permanent nightsides \citep[e.g.,][]{Bell18,Komacek18,Tan19}. In particular, molecules such as water are expected to dissociate on their dayside hemispheres \citep[e.g.,][]{Lothringer18}, many atomic elements are found in their ionized states, and temperature-pressure (T-P) profiles often exhibit thermal inversions. On the cooler nightsides, various atomic species may recombine or condense out of the atmosphere. Dynamical processes such as re-circulation, vertical mixing, and global heat transport, as well the unique physical properties of individual systems, dictate the extent to which chemistry on the nightside impacts chemistry on the dayside, and vice versa.

In recent years, high-resolution spectroscopy at optical wavelengths has proven to be a particularly powerful probe of UHJ atmospheres. The atomic, ionic, and molecular species expected to be present in their atmospheres exhibit hundreds of their strongest spectral features at these wavelengths, allowing for simultaneous detections of multiple species \citep[e.g.,][]{Ehrenreich20,Hoeijmakers22,Prinoth23,Pelletier23}. High-resolution optical spectroscopy can be used to study UHJs both during transit and throughout their pre- and post-eclipse orbital phases. Such observations offer us a global picture of their atmospheres, unveiling the distinct---yet likely coupled---chemical regimes that exist across these inherently 3D worlds.

With an ultra-hot equilibrium temperature $>$ 2600 K, a relatively short orbital period of 2.72 d, and a hot, bright host star (HD~133112; A61V--V; V = 6.6; $T_\mathrm{eff}$ $\sim$ 8000 K), the UHJ WASP-189b \citep{Anderson18} is particularly amenable to atmospheric characterization. Since its discovery in 2018, WASP-189b has been studied in great detail at both high- and low-resolution, and via both transmission and emission spectroscopy. 
Although early attempts to characterize its atmosphere resulted in null detections \citep{Cauley20}, the atmosphere of WASP-189b was eventually detected by \cite{Yan20}, who used high-resolution dayside observations from the High Accuracy Radial velocity Planet Searcher in the Northern hemisphere (HARPS-N) at the Telescopio Nazionale \textit{Galileo} (TNG) to observe Fe emission lines indicative of a thermal inversion. Soon after, \cite{Lendl20} used the CHaracterising ExOPlanets Satellite (CHEOPS) to observe four occultations and two transits of WASP-189b, leading to refined measurements of various planetary parameters. \cite{Deline22} later combined these data with full phase curve observations of WASP-189b from CHEOPS, further refining the system parameters and reporting no significant hotspot offset from the phase curve.

Using high-resolution transmission spectroscopy from both HARPS on the European Southern Observatory (ESO) 3.6 m telescope at La Silla Observatory and HARPS-N at the TNG, \cite{Prinoth22} reported significant detections of 9 species (Fe, Cr, Mg, Mn, Ti, V, Fe${}^+$, Ti${}^+$ and TiO) in the atmosphere of WASP-189b. These results represented the first unambiguous detection of TiO in an exoplanet's transmission spectrum.
Through analyzing the differing line positions of these detected species, \cite{Prinoth22} further explored the three-dimensional thermo-chemical stratification present in the planet's atmosphere and concluded that the different species originated in different thermal, chemical, and/or dynamical regimes. \cite{Langeveld22} analyzed these same HARPS and HARPS-N observations and reported a detection of Na, which was tentatively detected by \cite{Prinoth22}. Likewise, \cite{Stangret22} analyzed a subset of these data and reported detections of Fe, Fe${}^+$, and Ti. \cite{Gandhi23} analyzed these observations in a retrieval framework and reported chemical abundances of eleven neutral atomic and molecular species (Fe, Mg, Ni, Cr, Mn, V, Ca, Ti, TiO, TiH, and Na; note that in the case of Na a 2$\sigma$ upper limit was reported). \cite{Lee22} used a 3D general circulation model (GCM) to further study these detections.

More recently, \cite{Prinoth23} combined these HARPS and HARPS-N observations with additional transit data from the Echelle SPectrograph for Rocky Exoplanets and Stable Spectroscopic Observations (ESPRESSO) at the Very Large Telescope (VLT) and MAROON-X at Gemini North. Using a similar analysis as in \cite{Prinoth22}, \cite{Prinoth23} were able to detect H, Na, Mg, Ca, Ca${}^+$, Ti, Ti${}^+$, TiO, V, Cr, Mn, Fe, Fe${}^+$, Ni, Sr, Sr${}^+$, and Ba${}^+$ in the atmosphere of WASP-189b via time-resolved spectroscopy, of which Sr, Sr${}^+$, and Ba${}^+$ were new detections. \cite{Prinoth23} also showed that while the majority of these species exhibit an increase in signal strength over the course of the transit (likely due to the larger atmospheric scale height of the hotter trailing terminator), several species exhibit a constant or decreasing signal strength. Notably, the signal strength of TiO decreases over the course of the transit, while the signal strength of Ti remains roughly constant. The authors hypothesize that TiO largely dissociates on the hotter dayside \citep[e.g.,][]{Cont21}, while Ti partly ionizes to Ti${}^+$ \citep{Prinoth23}. Finally, \cite{Prinoth24} analyzed particularly strong absorption lines in their MAROON-X observations through narrow-band spectroscopy.

In the infrared, \cite{Yan22} used the GIANO-B spectrograph at the TNG to detect CO in the dayside atmosphere of WASP-189b. As with the Fe lines detected in \cite{Yan20}, the CO signal appeared in emission, indicating the presence of a thermal inversion in the atmosphere. In the near-ultraviolet (NUV), \cite{Sreejith23} used three transit observations from the Colorado Ultraviolet Transit Experiment (\textit{CUTE}) CubeSat to detect escaping metals (Mg${}^+$ and possibly Fe${}^+$) in the planet's upper atmosphere.

In this paper, we present high-resolution optical spectroscopy of WASP-189b's dayside atmosphere obtained with the new Gemini High-Resolution Optical SpecTrograph (GHOST; \citealt{Ireland12,Ireland14,GHOST22,McConnachie22,McConnachie24}; Kalari et al.~2024, submitted) at the Gemini South Observatory in Chile. The observations were obtained as part of the instrument's System Verification\footnote{\url{https://www.gemini.edu/instrumentation/ghost/ghost-system-verification}} (SV) run in May 2023. Our aim is to assess GHOST's potential for high-resolution characterization of exoplanet atmospheres by comparing these observations to past work with similar instruments, as well as to characterize WASP-189b's atmosphere with our GHOST data. We detect the planet's atmosphere at a high significance and recover the neutral iron emission feature detected by \cite{Yan20}, validating the presence of a thermal inversion in WASP-189b's atmosphere. We also search for additional atmospheric species, and discuss the implications of model injection/recovery tests on the planet's atmospheric composition.

This paper will proceed as follows. In Section \ref{sec:obs}, we introduce GHOST at Gemini South and describe the WASP-189b observations obtained during the instrument's SV run. We outline our data reduction methods in Section \ref{sec:reduc}, including a brief overview of the GHOST data reduction software. Our analysis and modelling methods are presented in Section \ref{sec:analysis}, and our results and a discussion follow in Section \ref{sec:results}. We conclude in Section \ref{sec:conclusion}, and an Appendix follows.

\section{Observations}
\label{sec:obs}
We obtained one phase curve observation of WASP-189b over the course of approximately 3 hours with GHOST at the Gemini South Observatory in Chile. GHOST is a new facility instrument at Gemini South, available for regular queue observations starting in the 2024A observing semester. The observations described herein were obtained as part of the SV observing run that took place in May 2023, in order to prepare the instrument for general community use. WASP-189b was chosen as a target due to previous high-significance detections of its atmosphere \citep[e.g.,][]{Yan20}, offering an excellent comparison for GHOST, as well as the ease of scheduling observations during the relatively short ($\sim$1 week) SV period. The raw and reduced data products are available in the Gemini Observatory Archive (GOA) under the program ID GS-2023A-SV-102\footnote{\url{https://archive.gemini.edu/searchform/GS-2023A-SV-102-10}} (PI: Deibert).

Below we provide a brief introduction to GHOST, followed by details about our WASP-189b phase curve observations.

\subsection{The Gemini High-Resolution Optical SpecTrograph (GHOST)}
\label{subsec:ghost}
GHOST is a new facility-class high-resolution optical spectrograph available at the Gemini South Observatory. The instrument has the capability to obtain spectra from 347 to 1060 nm in a single exposure (with the useful range being 383 to 1000 nm; Kalari et al.~2024, submitted). 
Spectra can be obtained in either a ``standard resolution'' mode with a resolving power of $R \sim$ 56,000 (with the option to observe two targets simultaneously) or a ``high-resolution'' mode with a resolving power of $R \sim$ 76,000. The instrument is composed of both a blue and a red arm, with light being split at 530 nm by a dichroic. Separate exposure times and readout modes can be set for the blue and red arms, however we did not do so for this analysis (see Section \ref{subsec:obs}). As the blue and red arms consist of different cameras and detectors, we carried out our analysis separately on the blue and red throughout this work. Further details on the instrument design, available modes, and early scientific performance can be found in \cite{McConnachie24}, Kalari et al.~(2024, submitted) and references therein, as well as the GHOST web pages{\footnote{\url{https://www.gemini.edu/instrumentation/ghost}}.

\subsection{WASP-189b Dayside Observations}
\label{subsec:obs}
We observed the phase curve of WASP-189b for approximately three hours on May 13th, 2023 (UT) using the high-resolution mode of GHOST. As the target itself was a bright (V = 6.6) point source, we set the the instrument's binning mode to 1 x 4 (spectral x spatial) and the read mode for both the blue and red detectors to ``medium''. This resulted in a combined readout and write time of $\sim$21 seconds per spectrum. The instrument was set to the high-resolution single-target mode, with the science IFU centered on the target and a dedicated sky IFU observing the sky.

\begin{figure}
    \centering
    \includegraphics{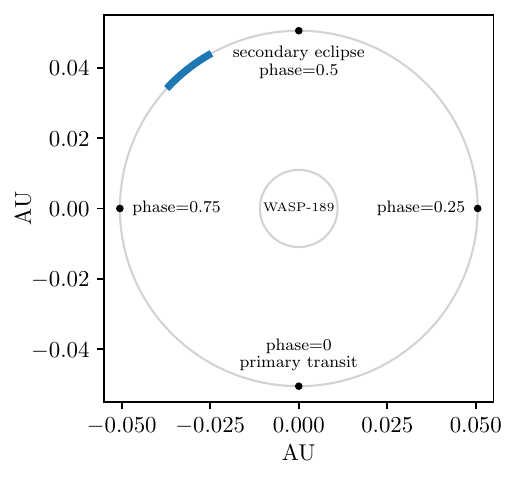}
    \caption{A visualization of the post-eclipse orbital phases ($\sim$0.58 to $\sim$0.63) covered by these observations. The planetary orbit and host star are drawn to-scale. Note that the brief gap in our observations is not visible, but can be seen in Fig.~\ref{fig:airmass}.}
    \label{fig:phases}
\end{figure}

The observations are summarized in Table \ref{tab:obs}. In total, we obtained 165 spectra covering orbital phases of $\sim$0.58 to $\sim$0.63 (i.e., following secondary eclipse; see Fig.~\ref{fig:phases}), which we determined using the orbital parameters in Table \ref{tab:params}. Of these 165 spectra, the first spectrum was obtained with an exposure time of 60~s; however the observer on duty noted that the spectrum was saturated and reduced the exposure time to 45~s for the remainder of the observing period, resulting in a $\sim$66~s cadence when accounting for the readout and write times. The first over-exposed spectrum was discarded from our analysis. Additionally, two files (occurring approximately a quarter and halfway through the observations) could not be processed by the data reduction software and were not included in our analysis. In total, our analysis therefore made use of 162 45-s exposures of the phase curve of WASP-189b. 

The GHOST control software crashed for approximately 20 minutes during the observing period, leading to a small gap in our phase coverage from 0.591 to 0.597. This was due to software issues while the instrument was still being integrated into Gemini's regular observing queue, and is unlikely to occur in future observations. Given that our observations are not as time-critical as an exoplanet transit, however, this did not impact our analysis, nor did it affect the exposures that were obtained. The total observing period was therefore $\sim$3.3 hours, with $\sim$3 hours of this time spent on-source and $\sim$0.3 hours lost to technical difficulties.

\begin{deluxetable*}{c c c c c c}
\label{tab:obs}
\tablecaption{Summary of GHOST/Gemini South observations of WASP-189b used in this analysis.}
\tablehead{%
    \colhead{Date (UT)} & \colhead{Num.~Exposures} & \colhead{Exposure Time (s)} & \colhead{Orbital Phase Coverage} & \colhead{Airmass Variation} & \colhead{{SNR Variation\tablenotemark{a}}}
    }
\startdata
13 May 2023 & 162 & 45 & 0.58 -- 0.63 & 1.12 -- 1.53 & {191 -- 233 (Blue)} \\
 & & & & & {203 -- 243 (Red)} \\
\enddata
\tablenotetext{a}{The quoted SNR variation includes all orders.}
\end{deluxetable*}

The airmass varied from $\sim$1.12 to $\sim$1.53 throughout the observations, as shown in Fig.~\ref{fig:airmass}. There was thin cirrus cloud cover present throughout the night, and the seeing fell within the 70th and 85th percentile bins typical of Cerro Pach\'{o}n (i.e., between $\sim$0.75 and 1.05 arcsec). 

\begin{figure}
    \centering
    \includegraphics{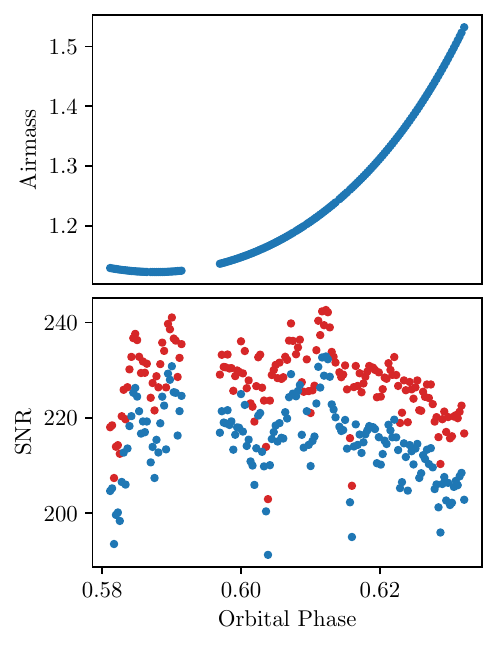}
    \caption{The change in airmass {(top panel) and SNR (bottom panel)} throughout the observations. {The SNR is plotted separately for the blue and red arms of the instrument, and is the average SNR per pixel per exposure across all orders. The SNR appears to decrease with increasing airmass at the end of our observations, as expected.} The $\sim$20 minute gap in our observing coverage is visible.}
    \label{fig:airmass}
\end{figure}

The average signal-to-noise ratio (SNR) of all obtained spectra over the full GHOST wavelength range was $\sim$215 per pixel in the blue and $\sim$228 per pixel in the red, though we note that there is a sharp drop-off in the instrument's throughput and sensitivity at edge orders (Kalari et al.~2024, submitted)\footnote{See also \url{https://www.gemini.edu/instrumentation/ghost/capabilities}}. {The SNR remained roughly constant throughout the observations, varying from $\sim$191 to $\sim$233 on average in the blue and $\sim$203 to $\sim$243 on average in the red. This variation seems to trend with airmass; as seen in Fig.~\ref{fig:airmass}, the SNR decreases slightly towards the end of the observations as the airmass begins to increase.}
For our analysis, we chose to exclude exclude orders with an average SNR $<$ 100, which resulted in the removal of 12 orders at the blue end of the spectra, 2 orders at the red end of the spectra, and 2 orders around $\sim$530 nm (corresponding to the edges of the blue and red detectors). This threshold was somewhat arbitrary, though informed by the fact that the drop-off in sensitivity at edge orders corresponded to orders with SNR $\lesssim$ 100 in our data; and also informed by recommendations presented in \cite{BoldtChristmas23}. Fig.~\ref{fig:raw-spectrum} provides a visualization of where these excluded orders occurred.

\begin{deluxetable*}{l c c c}
\label{tab:params}
\tablecaption{Planetary system parameters used in this work.}
\tablehead{%
    \colhead{Parameter} & \colhead{Symbol [unit]} & \colhead{Value} & \colhead{Reference}
    }
\startdata
Transit Midpoint & $T_0$ [BJD] & 2456706.4566 $\pm$ 0.0023 & \cite{Ivshina22} \\
Period & p [days] & 2.7240308 $\pm$ 0.0000028 & \cite{Ivshina22} \\
Planetary Radius & $R_p$ [$R_J$] & 1.600${}^{+0.017}_{-0.016}$ & \cite{Deline22} \\
Planetary Mass & $M_p$ [$M_J$] & 1.99${}^{+0.16}_{-0.14}$ & \cite{Lendl20} \\
Planetary Equilibrium Temperature & $T_\mathrm{eq}$ [K] & $>2600$ & \cite{Lendl20} \\
Host Star Spectral Type & -- & A6IV--V & \cite{Anderson18} \\
Stellar Radius & $R_*$ [$R_\odot$] & 2.360 $\pm$ 0.030 & \cite{Lendl20} \\
Stellar Effective {Temperature} & $T_*$ [K] & 8000 $\pm$ 80 & \cite{Lendl20} \\
{Systemic Velocity} & {$\mathrm{RV}_\mathrm{sys}$ [km/s]} & {$-20.82 \pm 0.07$} & \cite{Yan20} \\
T-P Profile Temperature Point 1 & $T_1$ [K] &  $4320^{+120}_{-100}$ & \cite{Yan20} \\
T-P Profile Pressure Point 1 & $\mathrm{log}_{10}(P_1)$ [log bar] & $-3.10^{+0.23}_{-0.25}$ & \cite{Yan20} \\
T-P Profile Temperature Point 2 & $T_2$ [K] & $2200^{+1000}_{-800}$ & \cite{Yan20} \\
T-P Profile Pressure Point 2 & $\mathrm{log}_{10}(P_2)$ [log bar] & $-1.7^{+0.8}_{-0.5}$ & \cite{Yan20} \\
\enddata
\end{deluxetable*}

\section{Data Reduction}
\label{sec:reduc}

\subsection{Initial Reduction with \texttt{DRAGONS}}
We reduced the observations using version 1.0.0 of the GHOST Data Reduction software \citep{Ireland18,Hayes22}, which is a Python-based software utilizing the Gemini Observatory's \texttt{DRAGONS} data reduction platform \citep{dragons19,dragons22}{\footnote{\url{https://www.gemini.edu/observing/phase-iii/reducing-data/dragons-data-reduction-software}}. Briefly, this software performs a bias subtraction, flat-fielding, a cosmic ray rejection routine, and bad pixel masking, followed by an optimal extraction, wavelength calibration, and optional barycentric correction. The software outputs two data products from each arm: a 2D image with separate spectra from the individual orders, and a 1D spectrum where the orders have been combined (both including an extension containing the variance). We ran the reduction software with both the barycentric correction and the optional sky subtraction turned off, as we needed the data in the telluric frame for our telluric correction routine (see Section \ref{subsec:tellurics}) and from a visual inspection of the reduced sky spectra we determined that a sky subtraction was not necessary. The reduced data products as described above (along with a version where the barycentric correction and sky subtraction were turned on) are available in the GOA under the program ID GS-2023A-SV-102. An example spectrum reduced by the GHOST Data Reduction software, but prior to the additional corrections described below, is displayed in Fig.~\ref{fig:raw-spectrum}.

\begin{figure*}
    \centering
    \includegraphics[width=\textwidth]{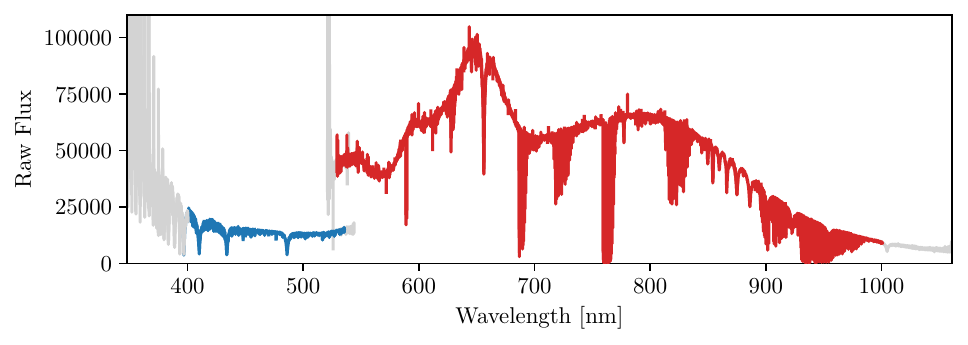}
    \caption{An example GHOST spectrum after reduction by the GHOST Data Reduction software but prior to any additional reduction undertaken for our analyses. The blue and red arms of the data are coloured in blue and red, with regions excluded from our analysis (see Section \ref{subsec:obs}) coloured in grey. Although edge orders are excluded from both detectors near 530 nm, the overlap in the detectors' wavelength coverage means that we still have continuous coverage in our analysis.}
    \label{fig:raw-spectrum}
\end{figure*}

Following the extraction of the spectra from the raw images by the data reduction software, we carried out an additional outlier-flagging routine in order to correct any remaining cosmic rays or outliers. We replaced any points that deviated by more than 5 median absolute deviations with \texttt{nan} values (as in, e.g., \citealt{Deibert21}). Each spectrum was then median-normalized to account for varying flux levels.

\subsection{Telluric and Stellar Absorption Correction}
\label{subsec:tellurics}

After the initial data processing steps described above, the data are dominated by both stellar and telluric absorption lines, with the much weaker planetary signal buried by these stronger features. As the host star is a hot A-type star, tellurics are the dominant feature at this stage, with fewer stellar absorption features expected. As an initial correction, we created a median spectrum and divided this out of every individual spectrum. We then removed the remaining stellar and telluric features with the \textsc{SysRem} algorithm \citep{Tamuz05}, which is a principal component analysis (PCA)-like algorithm commonly used for this purpose \citep[e.g.,][among others]{Deibert21,Herman22,Deibert23}. Briefly, \textsc{SysRem} removes stationary features (in this case, the tellurics and the stellar absorption features) while {in theory} leaving the planetary signal---which is significantly Doppler-shifted over the course of the observations---intact. {However, we note that too many iterations of the \textsc{SysRem} algorithm can eventually begin to remove the planetary signal as well. This is evident in Fig.~\ref{fig:deltaCCF}, where more than $\sim$8 iterations of the algorithm begins to decrease the strength of a retrieved cross-correlation signal, likely due to the fact that the planetary signal is being removed. Furthermore, it has been demonstrated in recent works that \textsc{SysRem} can in fact cause a slight distortion of the planetary signal \cite[e.g.,][]{Gibson22}. This effect is not expected to have a signficant outcome on our results, but is important to consider when carrying out atmospheric retrievals.}

We ran the \textsc{SysRem} algorithm on the red and blue arms of the data in the telluric reference frame, using the 2D data product from the GHOST data reduction software without the barycentric correction applied. While the stellar features do have a slight Doppler shift, this shift is very small and they are essentially stationary in wavelength for the purposes of the \textsc{SysRem} algorithm. We used the airmass throughout the night as our initial guess of the first systematic to be removed, then ran between 1 and 20 iterations of the algorithm order-by-order on both the blue and red arms of the data. To determine the optimum number of iterations to apply to our data, we used the $\Delta$CCF method described in e.g., \cite{Spring22,Cheverall23}. Briefly, this method involves choosing the number of \textsc{SysRem} iterations which maximizes the peak SNR of the quantity $\Delta$CCF = CCF${}_\mathrm{inj}$ - CCF${}_\mathrm{obs}$, where CCF${}_\mathrm{inj}$ and CCF${}_\mathrm{obs}$ are the cross-correlation functions (CCFs) calculated via the Doppler cross-correlation technique (see Section \ref{subsec:xcorr}) for a model-injected version of the data and the observed data, respectively. To calculate the SNR of the $\Delta$CCF map, the signal comes from the $\Delta$CCF map directly whereas the noise comes from CCF${}_\mathrm{obs}$. 

As discussed in \cite{Cheverall23}, this method is robust against biases that may come from optimizing \textsc{SysRem} on CCF${}_\mathrm{inj}$ or CCF${}_\mathrm{obs}$ individually. This method is also largely insensitive to different models or injection velocities \citep{Cheverall23}. \cite{Cheverall23} also suggested that using a wide range of velocities in the CCFs can limit biases in calculating the SNR.

While this method can in principle be applied on an order-by-order basis, optimizing each order for a different number of \textsc{SysRem} iterations, we instead carried out a global $\Delta$CCF optimization for the red and blue detectors separately and leave an in-depth exploration of the robustness of an order-by-order strategy for future work. To do this, we injected an Fe model (see Section \ref{subsec:models}) into the data at the expected planetary velocity, and repeated our reduction and analysis process on this model-injected data for \textsc{SysRem} iterations ranging between 1 and 20. We chose Fe as it was the only species detected at high significance (see Section \ref{sec:results}), though we reiterate that \cite{Cheverall23} demonstrated that the $\Delta$CCF method shows consistency between different models. 

The results of this test are shown in Fig.~\ref{fig:deltaCCF}. In the red arm, we found that the $\Delta$CCF detection significance peaked after 6 iterations of the \textsc{SysRem} algorithm, while in the blue, this peak occurred after 8 iterations. We therefore adopt 6 and 8 as the optimum number of iterations for the red and blue arms, respectively, and use these throughout the rest of our analysis. 

\begin{figure}
    \centering
    \includegraphics{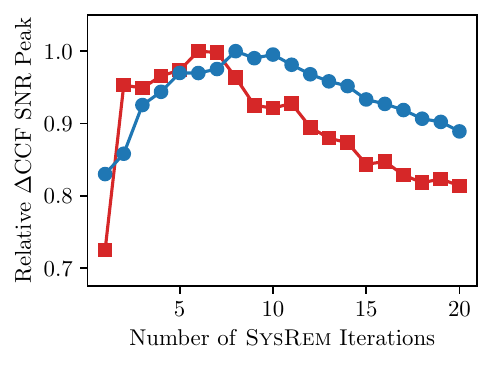}
    \caption{The peak SNR of the $\Delta$CCF map for \textsc{SysRem} iterations varying from 1 to 20. The results for the red arm are indicated by red squares, while the results for the blue arm are indicated by blue circles. The values have been normalized to the maximum SNR of all iterations for each arm. We found that the peak SNR of the $\Delta$CCF map corresponded to 6 \textsc{SysRem} iterations in the red and 8 \textsc{SysRem} iterations in the blue. Beyond these values, the SNR steadily decreases, likely because the algorithm has begun to remove the planetary signal.}
    \label{fig:deltaCCF}
\end{figure}

The results of applying the \textsc{SysRem} algorithm to our data are displayed in Appendix \ref{app:sysrem}. While we found that \textsc{SysRem} was sufficient in correcting tellurics and stellar absorption features for these observations, we aim to investigate additional telluric correction methods (including, e.g., \texttt{molecfit} or other methods which make use of synthetic telluric spectra) {as well as masking of particularly contaminted regions} in a future work. {Note that there is still some variation in the noise levels of the \textsc{SysRem}-corrected orders, and in some cases residual features due to tellurics are still visible. While applying different numbers of iterations of the algorithm on a per-order basis may help to further suppress these tellurics, we have opted to avoid additional fine-tuning, which may bias the results towards a particular detection. As mentioned previously, it could be worth exploring the efficacy of an order-by-order analysis in a future work.}

{In carrying out our analysis (see Section \ref{subsec:xcorr}) we weighted each pixel by its standard deviation (i.e., the final row of the figures in Appendix \ref{app:sysrem}) following e.g., \cite{Snellen10,Esteves17}. This ensures that particularly contaminated regions of the data will not contribute excess noise.}

\section{Analysis}
\label{sec:analysis}

To characterize the atmosphere of WASP-189b, we carried out a Doppler cross-correlation analysis \citep[e.g.,][]{Snellen10} using high-resolution atmospheric models generated for WASP-189b's system parameters (see Table \ref{tab:params}). Below we describe the methods used to create these models and correlate them with our observations.

\subsection{Atmospheric Models}
\label{subsec:models}

We generated atmospheric models for a range of species using {version 2.7.7 of} \texttt{petitRADTRANS} \citep{petitradtrans}. We included species which (i) had high-resolution line lists available in \texttt{petitRADTRANS}, and (ii) had strong spectral lines present in the GHOST wavelength range. The system parameters used to generate these models are presented in Table \ref{tab:params}. For the T-P profile, we used the best-fit parameters from the analysis in \cite{Yan20}, who retrieved a two-point T-P profile from their HARPS-N observations following the parametrization of \cite{Brogi14}. Briefly, the temperature is assumed to be isothermal above/below the points ($T_1$, $P_1$) and ($T_2$, $P_2$), where $T$ refers to temperature and $P$ refers to pressure, and change linearly with $\mathrm{log}_{10}(P)$ between the two points with a gradient of:
\begin{equation}
    T_\mathrm{slope} = \frac{T_1 - T_2}{\log P_1 - \log P_2}.
\end{equation}

The best-fit parameters from \cite{Yan20}, which we used for this nominal T-P profile, are displayed in Table \ref{tab:params}. The resulting T-P profile is shown in the right-hand side of Fig.~\ref{fig:fe_model}.

As in \cite{Yan20}, our models were generated assuming a solar metallicity and neglecting H${}^-$ opacity and Rayleigh scattering. To determine the abundance of each species, we ran the \texttt{FastChem} chemical equilibrium model \citep{Stock18}, following e.g., \cite{Johnson23,Petz23}. We used the same assumptions as in our model spectra; i.e., {we assumed solar abundances and metallicities, and used} the best-fit two-point T-P profile from \cite{Yan20}. We use the volume mixing ratio (VMR) as a function of pressure output from \texttt{FastChem} when generating our models with \texttt{petitRADTRANS}. We note that this is slightly different from \cite{Yan20}, who assumed a constant VMR. We converted these \texttt{FastChem} abundances to mass fractions following the instructions in the \texttt{petitRADTRANS} tutorial. As in \cite{Yan20}, we scaled our models by the blackbody spectrum of the star (generated with \texttt{petitRADTRANS} using the stellar parameters displayed in Table \ref{tab:params}) to obtain the model spectrum in the form 1 + $F_p$/$F_*$, where $F_p$ is the planetary flux from \texttt{petitRADTRANS} and $F_*$ is the stellar blackbody. When cross-correlating the models with our observations, we subtracted off the continuum following \cite{Herman22}.

We generated individual models in this way for Fe, Fe${}^+$, Ti, V, V${}^+$, Al, Ca, Ca${}^+$, Cr, K, Mg, Na, Si, TiO, VO, FeH, and OH. We also generated one model which contained all of these species. 

Of these, Fe, Fe${}^+$, Ti, V, V${}^+$, Al, Ca, Ca${}^+$, Cr, Mg, and Si were contributed to \texttt{petitRADTRANS} by K. Molaverdikhani from the Kurucz line lists\footnote{\url{http://kurucz.harvard.edu/}}. For TiO we made use of the ${}^{48}$TiO line list sourced from Exomol \citep{McKemmish19}; VO was sourced from Exomol \citep{McKemmish16}; FeH was sourced from Exomol \citep[see references in][]{petitradtrans}; OH was sourced from HITEMP \citep[see references in][]{petitradtrans}; and K and Na were sourced from VALD with Allard wings \citep[see references in][]{petitradtrans}. {We note that line lists provided in \texttt{petitRADTRANS} are not necessarily in the same reference frame. To match the data, we therefore converted all wavelengths to air wavelengths where necessary. Table \ref{tab:linelists} in Appendix \ref{app:linelists} details which line lists required this conversion.}
An example Fe model, along with the associated T-P profile used to generate the model, is shown in Fig.~\ref{fig:fe_model}. 

As mentioned previously, these are species which have strong spectral lines in the GHOST wavelength range and which have high-resolution line lists available in \texttt{petitRADTRANS}. However, because we were also interested in titanium chemistry in the atmosphere (see the discussion on \citealt{Prinoth23} in Section \ref{sec:intro}), we generated an additional binary mask for Ti${}^+$, which is not available in the collection of high-resolution line lists offered by \texttt{petitRADTRANS}. For this, we made use of the NIST 2014-09-16 database sourced from \texttt{Cloudy} \citep{cloudy}.

\begin{figure*}
    \centering
    \includegraphics{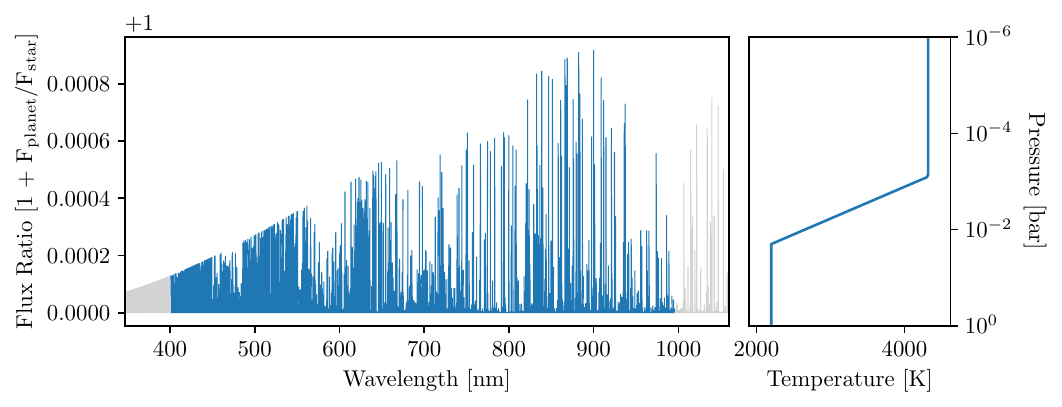}
    \caption{{\textit{Left:}} The Fe model used in our analysis, generated with \texttt{petitRADTRANS} \citep{petitradtrans} as described in Section \ref{subsec:models}. The full GHOST wavelength range is displayed; however, we note that we only used a subset of this wavelength range in our analysis, indicated by the blue region (see Section \ref{subsec:obs}). {\textit{Right:}} The two-point T-P profile used to generate the model, using the best-fit parameters from \cite{Yan20} and the planetary parameters listed in Table \ref{tab:params}.}
    \label{fig:fe_model}
\end{figure*}

\subsection{Doppler Cross-Correlation}
\label{subsec:xcorr}
Following the initial data reduction steps described in Section \ref{sec:reduc}, we analyzed the data through the Doppler cross-correlation technique \citep[e.g.,][]{Snellen10,Deibert23}. We shifted the data to the stellar rest frame using the barycentric radial velocity calculated by the data reduction software {and the systemic velocity from \citealt{Yan20} ($\mathrm{RV}_\mathrm{sys} = -20.82 \pm 0.07$ km/s; see Table \ref{tab:params})}. The model spectra were convolved with an average, non-varying instrumental profile which we calculated following \cite{Herman22} and using the \texttt{convolve} and \texttt{Gaussian1DKernel} functions from \texttt{Astropy} \citep{astropy:2013, astropy:2018, astropy:2022}. We then Doppler-shifted the model spectra to radial velocities (RV) ranging from -300 km/s to +300 km/s with step-sizes of 1 km/s and cross-correlated the models with our observations at every velocity to create cross-correlation functions (CCFs). While the orbital velocity trace of the planet may in some cases be strong enough to be seen by eye at this step, we further phase-folded the data to a range of Keplerian orbital velocities ($K_p$) ranging from 1 km/s to 300 km/s with step-sizes of 1 km/s to create 2D $K_p$-RV maps for each species. At this point, an atmospheric detection should be visible as a peak in the 2D $K_p$-RV map. As the data have been shifted to the stellar rest frame, we expect this signal to occur at RV~$=$~0 km/s and $K_p \sim 197$ km/s (from Kepler's third law and the parameters in Table \ref{tab:params}).

As discussed previously, we carried out the analysis separately for the blue and red detectors and only combined the two arms of the data at the very end (see Section \ref{sec:results}).

\subsection{Detection Significances}
\label{subsec:detsig}

Recent work has shown that there can be discrepancies between detection significances calculated via different methods \citep[e.g.,][]{Cabot19,Spring22,Cheverall23}. We therefore used several different methods to determine our detection significances, and in general report the most conservative of these as our final results. First, we followed the standard methodology of dividing out the standard deviation of the 2D $K_p$-RV map. \cite{Cheverall23} demonstrated that this method is more robust if a sufficiently wide range of velocities are explored in the correlations, so we used a wide range in both $K_p$ and RV as explained in Section \ref{subsec:xcorr}. Similar to \cite{Smith24}, we calculated this standard deviation from a $\sigma$-clipped version of the map, where we take 5$\sigma$ as our clipping threshold (noting that a 3$\sigma$-clipped version yielded higher estimated significances). This is similar to methods which calculate the standard deviation excluding points located outside a window around the detection peak \citep[e.g.,][]{Deibert23}, though likely more robust as it does not rely on an exclusion window chosen by eye. 

Second, we followed e.g., \cite{Brogi13, Birkby13, BelloArufe22} in performing a Welch's \textit{t}-test with the \texttt{scipy.stats} package. As in \cite{BelloArufe22}, we split the cross-correlation values into ``in-trail'' and ``out-of-trail'' populations, where the ``in-trail'' population corresponded to the three pixels in each CCF closest to the expected planetary velocity, and the ``out-of-trail'' population corresponded to all other CCF values. While we calculated this value for all species, we note that \cite{Cabot19} suggested that the Welch's \textit{t}-test  may overestimate detection significances. This was indeed the case in our analysis, as seen in the following section.

A number of additional methods have been used to estimate detection significances in other works, including a ``bootstrapping'' method whereby the frames are randomly phase-scrambled and the analysis process is repeated N times \citep[e.g.,][]{Herman20,Herman22}. Similarly, recent work has shown that a log-likelihood map (which can be computed via a CCF-to-$\log L$ mapping; \citealt{Brogi19}) can be a more robust method for determining the detection significance, and additionally provides the advantage of being able to compare different models in a statistical framework \citep[e.g.,][]{Smith24}. We aim to explore these methods in a future work, but for the present we report the more conservative of our significances, and caution that these SNR values depend on the specifics of our calculations and therefore may contain some biases.

Informed by recent high-resolution cross-correlation analyses of exoplanet atmospheres, and by the fact that noise in the data can sometimes present itself as spurious $\sim$3$\sigma$ correlation peaks \citep[e.g.,][]{Esteves17,Cheverall23}, we take 5$\sigma$ as the detection threshold for the current work. 

\section{Results and Discussion}
\label{sec:results}

Following the methodology described in Section \ref{sec:analysis}, we cross-correlated our observations with model atmospheres generated for Fe, Fe${}^+$, Ti, V, V${}^+$, Al, Ca, Ca${}^+$, Cr, K, Mg, Na, Si, TiO, VO, FeH, and OH, with the all-species models. We also cross-correlated our data with a binary mask generated for Ti${}^+$. Our results are presented in the following sections.

\subsection{All-Species Model}
The results of cross-correlating our data with the all-species model are presented in Fig.~\ref{fig:full_result}. In the red arm, we detect the atmosphere at a significance of 7.9$\sigma$ via the standard deviation map and 10.3$\sigma$ via the Welch's \textit{t}-test, in line with the observation by \cite{Cabot19} that the Welch's \textit{t}-test may overestimate the detection significance. Interestingly, in the blue arm, the all-species model is only detected at a significance of 3.1$\sigma$ via the standard deviation map and 3.7$\sigma$ via the Welch's \textit{t}-test. We note that the spectral lines in the blue are generally weaker than those in the red for the species explored in our analysis, and that our blue-arm data had a lower average SNR than our red-arm data. Future GHOST observing strategies may benefit from separately adjusting the exposure times in the blue and red.  

\begin{figure*}
    \centering
    \includegraphics[width=\textwidth]{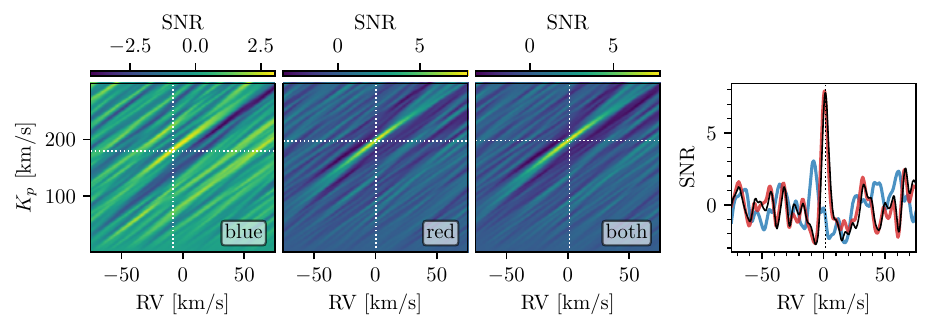}
    \caption{The results of cross-correlating our observations with the all-species model. {\textit{Left three panels:}} The 2D $K_p$-RV maps for the blue, red, and combined results, respectively, with the location of the peak correlation signal indicated in each case by white dashed lines. The colorbars indicate the significances of the maps, determined via dividing out the standard deviation away from the peak signal (see Section \ref{subsec:detsig}). A correlation signal is visible near the expected planetary location in all cases, though significantly stronger in the red than the blue. {\textit{Right:}} Slices of the 2D $K_p$-RV maps at the peak $K_p$ for the blue (blue line), red (red line), and combined (black line) results. The SNR is the same as that indicated by the colorbars. The dashed black line indicates the peak RV of the combined results.}
    \label{fig:full_result}
\end{figure*}

In combining the blue and red results, we weight each detector by the expected flux of the atmospheric model in the detector's wavelength range. For most models, this down-weights the blue arm, as most species we investigated contain weaker spectral lines in this region. In the case of the all-species model, this combination yields a detection of 7.8$\sigma$ via the standard deviation ({11.5}$\sigma$ via the Welch's \textit{t}-text). These results suggest that the majority of the detected signal is coming from the red arm of the data (and as we explain in the following section, the majority of this detection is likely due to Fe lines in the atmosphere).

In the combined map, the peak significance is detected at a Keplerian orbital velocity of $K_p = 198^{+17}_{-22}$ km/s and a radial velocity of RV = $1.4_{-11.3}^{+12.0}$ km/s, where we have taken as error the 1$\sigma$ extent of the correlation peak (as in, e.g., \citealt{Deibert23}). These velocities are consistent with previous analyses of WASP-189b \citep[e.g.,][]{Yan20}, which we discuss in more detail in the following section.

\subsection{Neutral Iron Detection}

The results of cross-correlating our data with the atmospheric model generated for Fe (i.e., Fig.~\ref{fig:fe_model}) are presented in Fig.~\ref{fig:fe_result}. In the red, we detect Fe at a significance of {7.5}$\sigma$ via the standard deviation map (10.7$\sigma$ via the Welch's \textit{t}-text). Interestingly, in the blue arm alone, we do not recover a significant detection of Fe. 

The combination of the blue and red data yields a detection significance of 5.5$\sigma$ via the standard deviation map ({8.8}$\sigma$ via the Welch's \textit{t}-test). In other words, adding the blue data to the red data in the case of our Fe CCFs only serves to add noise and weaken our detected signal. A similar effect was noted by \cite{Herman22}, who detected Fe in the dayside atmosphere of WASP-33b but were only able to make a significant detection if the blue data (in their case, wavelengths shorter than $\sim$600 nm) were excluded from their analysis. They showed that this was due to the low line contrast of Fe in the blue (see, for e.g., Fig.~\ref{fig:fe_model}), which is likely also limiting our detection capabilities. 

As discussed in more detail in Section \ref{subsec:nondetec}, however, we found that we were able to recover an injected Fe model in the blue. Although we note that an injected model will naturally be recovered at a higher significance than the real atmosphere (as we are effectively correlating the model with itself), this could also indicate that our models are over-estimating the abundance of Fe. Future work employing more sophisticated modelling techniques and/or atmospheric retrievals may be able to shed light on this possibility. For the present work, we base our result (as shown in Fig.~\ref{fig:fe_result}) off the red-arm data only.

As can be seen in Fig.~\ref{fig:fe_result}, the detected signal is very broad; this is due to the limited phase coverage of our data \citep[see for e.g.,][]{Yan20}. Additional observations covering pre-eclipse orbital phases, or a combination of our data with the existing HARPS-N observations, would allow us to place more stringent constraints on the planet's velocity; however, this is outside the scope of the present work.

\begin{figure}
    \centering
    \includegraphics[width=\columnwidth]{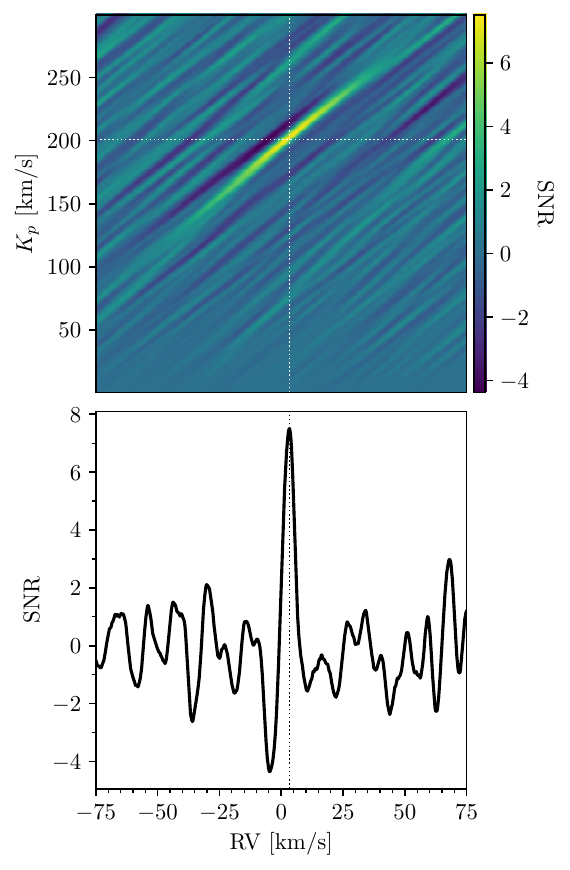}
    \caption{The results of cross-correlating our red-arm observations with an atmospheric Fe model (see Fig.~\ref{fig:fe_model}). {\textit{Top:}} The 2D $K_p$-RV map, with the location of the peak correlation signal indicated by white dashed lines. The colorbar indicates the significance of the map, as determined by dividing out the standard deviation of regions away from the peak signal (see Section \ref{subsec:detsig}). A clear correlation signal is visible near the expected planetary location. {\textit{Bottom:}} A slice of the 2D $K_p$-RV map at $K_p$ = 201 km/s (i.e., the $K_p$ at the peak correlation signal). The dashed black line indicates the location of the signal peak, and the SNR is the same as that indicated by the colorbar.}
    \label{fig:fe_result}
\end{figure}

As in \cite{Yan20}, the detected Fe lines are found in emission, indicating the presence of a thermal inversion in WASP-189b's atmosphere. This is consistent with theoretical predictions from e.g., \cite{Lothringer18, Lothringer19}, and with analyses of comparable UHJs which have also been shown to exhibit thermal inversions in their atmospheres \citep[e.g.,][among many others]{Pino20, Nugroho20, Kasper21}. While the iron in WASP-189b's atmosphere is likely at least partially responsible for the thermal inversion, additional observations will be needed to determine whether other strong optical absorbers (e.g., TiO) could also be contributing.

The peak correlation signal is located at a planetary orbital velocity of $K_p$ = $201^{+18}_{-17}$ km/s and a radial velocity of RV = $3.4^{+12.5}_{-9.0}$ km/s. These values are consistent within uncertainties with those expected from Kepler's laws and those determined by \cite{Yan20}, as well as those from the all-species model, though we note again that our values are poorly constrained due to the limited orbital phase coverage of our observations. In particular, our $K_p$ value is consistent with the \cite{Yan20} value of $K_p = 193.54^{+0.54}_{-0.57}$ km/s, and our RV is consistent with their value of RV = $0.66^{+0.25}_{-0.26}$ km/s (note that \cite{Yan20} also reported a redshifted RV). While this redshifted RV value could be due to winds or dynamical processes in the atmosphere, additional data at a wider range of orbital phases are needed to better constrain the exact value and determine whether it is indeed offset from zero. We note that this offset could also be due to errors in our transit ephemeris (e.g., Meziani et al., in prep.) {or errors in the systemic velocity used to shift to the planet rest frame. Indeed, we note that there are differences of a few km/s between the systemic velocities measured by \cite{Yan20}, \cite{Anderson18}, and \cite{Prinoth24}, though the values are generally within uncertainties of each other. Errors in these values could manifest as a few km/s offset in the final cross-correlation signal, and we thus caution against explicitly interpreting these offsets as winds and/or dynamics.}

\subsection{Non-Detections and Model Injection/Recovery Tests}
\label{subsec:nondetec}
The remaining species investigated in this work did not yield significant (i.e., $>$5$\sigma$) detections in the blue, red, or combined datasets. We present the combined blue and red 2D $K_p$-RV maps for these undetected species in Fig.~\ref{fig:nondetec}. 

\begin{figure*}
    \centering
    \includegraphics[width=\textwidth]{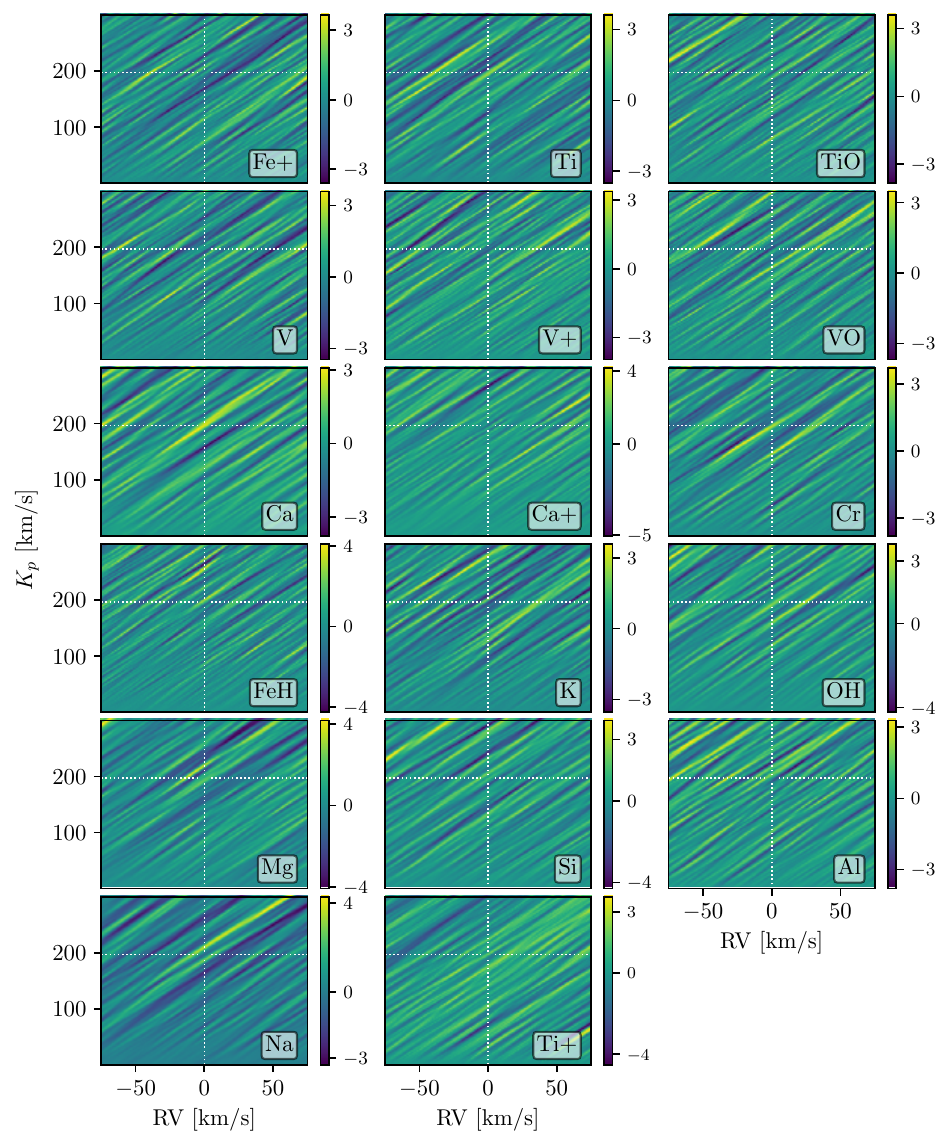}
    \caption{The combined blue and red 2D $K_p$-RV maps for the species which were not detected in this work. The plots are as described in the caption of Fig.~\ref{fig:fe_result}, and the species are indicated in the bottom-right corners. The expected location of the planetary signal (i.e., RV~=~0~km/s, $K_p$ $\sim$ 197 km/s) is marked in dashed white lines. The colorbar indicates the SNR as calculated via the standard deviation map.}
    \label{fig:nondetec}
\end{figure*}

Of these non-detections, we note that Ca and Mg exhibit peak correlations of $\sim$3$\sigma$ and $\sim$4$\sigma$ respectively near the expected planetary locations (though the Mg signal is offset by $\sim$-10~km/s). In the case of Ca, this signal may be worth following up with additional observations; however, previous studies have shown that residual noise can cause spurious $\sim$3$\sigma$ peaks \citep[e.g.,][]{Esteves17}. In the case of Mg, we note that \cite{Petz23} recently reported a tentative Mg signal in the atmosphere of the UHJ KELT-20b using observations from the Potsdam Echelle Polarimetric and Spectrographic Instrument (PEPSI) on the Large Binocular Telescope (LBT). Their observations spanned two bandpasses simultaneously, with wavelength coverage from 480 to 544.1 nm (blue) and 627.8 to 741.9 nm (red), though they only used the blue-arm data in their Mg analysis. Although they did report a cross-correlation signal at the expected location, they showed that this signal was likely caused by spurious correlations with Fe lines in the planet's atmosphere, which may also be the case here. As a test, we correlated our data with a model containing both Fe and Mg, as this might be expected to yield a higher detection significance than the Fe-only model if Mg were indeed present in the atmosphere. This correlation yielded a similar detection strength to that with the Fe-only model, suggesting that tentative Mg signal in our $K_p$-RV map may be due to a spurious correlation with the Fe lines or noise in the data.

Overall, these non-detections could indicate that these species are not present in the region of the atmosphere probed by our observations, or that our data are not sufficient to detect them if they are present. We do note that many of these species have previously been detected in the terminator region via transmission spectroscopy, albeit with more data (see the discussion in Section \ref{sec:intro}); however, we are probing a different region of the atmosphere with the present observations, and the signals themselves are expected to be weaker in emission (see for e.g., \citealt{Johnson23}). 

To investigate these non-detections in more detail, we carried out model injection/recovery tests for each individual species. We did this by injecting our models into the data at the negative of the expected planetary Keplerian velocity $K_p$, and repeating our reduction and analysis processes. We injected at $-K_p$, rather than $+K_p$, to avoid ``boosting'' any weak, undetected signals that may be present. Although Fe was detected in our fiducial analysis, we included it in the rejection/recovery tests as well in order to compare our real detection with that expected from our model (although we note that in the case of the model injection/recovery tests, we are effectively cross-correlating the model with itself). The results of these model injection/recovery tests are presented in Fig.~\ref{fig:injection}.

\begin{figure*}
    \centering
    \includegraphics[width=\textwidth]{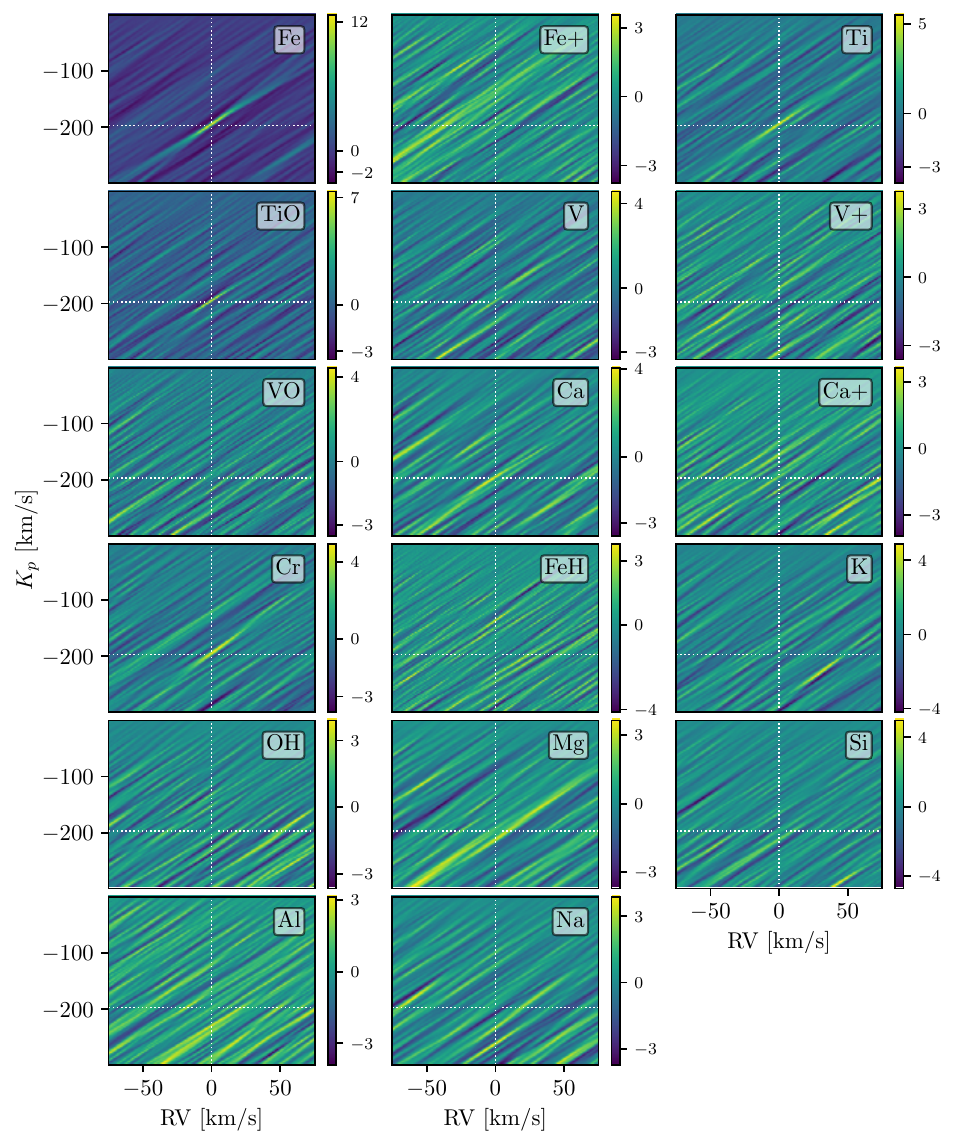}
    \caption{The results of our model injection/recovery tests, as described in Section \ref{subsec:nondetec}. The plots are as described in the caption of Fig.~\ref{fig:nondetec}, though we note that the injected signal is expected to be present at $-K_p$ rather than $K_p$ (as indicated by the white dashed lines).}
    \label{fig:injection}
\end{figure*}

We recovered the injected Fe model at a high significance of 12.6$\sigma$ (via the standard deviation) in the combined blue and red $K_p$-RV map, and individually at a significance of 12.8$\sigma$ in the red and 7.4$\sigma$ in the blue. Interestingly, these results suggest that we should have been able to detect Fe in the blue data (although the detection strength from the model injection/recovery test is naturally expected to be stronger than that of the real data). This could indicate that our model is over-estimating the abundance of Fe in the planet's atmosphere, or is otherwise not an ideal match to our data. Future work incorporating more sophisticated modelling techniques may be able to shed further light on these possibilities. 

Of the undetected species, the only injected models recovered at a significance of $>$5$\sigma$ were TiO and Ti. From the combined maps, we recover the TiO model at 7.4$\sigma$ and the Ti model at 5.5$\sigma$. We also tentatively recover the injected Ca model, at a significance of 4.0$\sigma$; the injected Cr model, at a significance of 4.9$\sigma$; and the Mg model, at a significance of 3.6$\sigma$. The remaining models did not yield strong signals at the expected location.

The results of our model injection/recovery tests indicate that if TiO and Ti were present in the dayside atmosphere of WASP-189b, we would have been able to detect them in our observations. Instead, as was also suggested by the transmission spectroscopy observations presented in \cite{Prinoth23}, it is likely the case that TiO has largely dissociated in the dayside atmosphere and Ti may have partially ionized to Ti${}^+$. Although we also did not detect Ti${}^+$ in the atmosphere, we note that that analysis was based off of a binary mask rather than a model, as Ti${}^+$ is not included in the high-resolution line lists available for download from \texttt{petitRADTRANS} (and therefore not included in our model injection/recovery tests). A more accurate model may lead to a detection, or may indicate that Ti${}^+$ is not detectable in the present data. Additional modelling and/or observations (or a combination of our data with the existing HARPS observations presented in \citealt{Yan20}) may help confirm these possibilities; however, this is beyond the scope of the present work.

Similarly, our model injection/recovery tests suggest that we may have marginally detected Ca if it were present in the regions of the atmosphere we're probing (albeit at a higher significance than the $\sim$3$\sigma$ feature we noted near the expected planetary location). At the same time, Ca may have ionized to Ca${}^+$, which we were not able to recover in our injection/recovery tests. Likewise, our model injection/recovery tests suggest marginal detections of Cr and Mg may have been possible. Although we did note a weak Mg signal at the expected location in our CCFs, this may have been complicated by the fact that the Mg model could produce an aliased detection with the Fe lines, as discussed previously \citep{Petz23}. Additional observations will be needed to determine whether Mg is present in the dayside atmosphere. 

\subsection{Comparison with Previous Work \& Future Prospects for Characterizing Exoplanet Atmospheres with GHOST}
As discussed in the introduction, the dayside atmosphere of WASP-189b has previously been observed at high-resolution in the optical with HARPS-N \citep{Yan20}. It has also been extensively studied with high-resolution transmission spectroscopy in the optical, using HARPS, HARPS-N, ESPRESSO, and MAROON-X \citep[e.g.,][]{Prinoth22,Langeveld22,Stangret22,Gandhi23,Prinoth23,Prinoth24}, as well as high-resolution dayside spectroscopy in the near-infrared using GIANO-B \citep{Yan22}.

Here we compare our results to the HARPS-N dayside spectroscopic observations, which probe a similar wavelength and orbital phase range (and thus similar region of the atmosphere) as our GHOST observations. In particular, HARPS-N has a high spectral resolution of R $\sim$ 115,000 over a wavelength range of 383 to 690 nm, allowing it to resolve hundreds of individual spectral lines in the optical. \cite{Yan20} used two nights of observations, covering orbital phases 0.533 -- 0.624 (post-eclipse) and 0.384 -- 0.497 (pre-eclipse), to detect Fe at 8.7$\sigma$ when both nights were combined. They also searched for, but were unable to detect, Fe${}^+$, Ti, Ti${}^+$, TiO, and VO. The SNR of their spectra ranged from 45 to 90 (with 60 s exposures) throughout the two nights of observations. As their observations sampled orbital phases both pre- and post-eclipse, they were able to place a tight constraint on the planet's Keplerian orbital velocity, finding $K_p = 193.54^{+0.54}_{-0.57}$ km/s \citep{Yan20}.

Our GHOST observations offer a comparable ({7.5}$\sigma$) detection in less observing time, covering only orbital phases from 0.58 -- 0.63 (resulting in a less well-constrained $K_p$). This demonstrates the advantages of Gemini's large 8-meter collecting area, offering us very high SNR spectra in relatively short exposures (45 s); as well as GHOST's short readout and write times, allowing us to obtain a large number of exposures and finely sample the planet's orbit. While GHOST has a lower resolution than HARPS-N, its very broad (383 to 1000 nm) wavelength coverage offers access to additional spectral lines, assuming tellurics at the redder end of the spectrum can be removed. In the case of Fe, this additional wavelength coverage contains many strong spectral lines (see Fig.~\ref{fig:fe_model}). Together, these properties make GHOST particularly well-suited to characterizing the daysides of hot exoplanet atmospheres at high spectral resolution in the optical.

Although the present observations do not offer a one-to-one comparison with other analyses of WASP-189b via transmission spectroscopy (as transmission spectroscopy probes a different region of the atmosphere), we note that our observations reach a comparable SNR to those from ESPRESSO and MAROON-X presented in \cite{Prinoth23,Prinoth24}, as demonstrated by Fig.~2 of the latter. We thus expect GHOST to be equally well-suited to characterizing exoplanet atmospheres via transmission spectroscopy in the optical, and to offer similar phase-resolved detections for particularly amenable targets.

\section{Conclusions}
\label{sec:conclusion}
In this work, we presented high-resolution optical spectroscopy of the dayside atmosphere of WASP-189b obtained with the new GHOST instrument at Gemini South. Using three hours of post-eclipse observations, we recovered a detection of neutral iron in the planet's atmosphere and verified the presence of a thermal inversion. We also carried out model injection/recovery tests for a wide range of atmospheric species, and discussed the implications of these tests on WASP-189b's atmospheric composition. Our results are consistent with a previous analysis of WASP-189b's dayside atmosphere in the optical \citep[][]{Yan20}, and with the trend among ultra-hot Jupiters to exhibit thermal inversions in their dayside atmospheres.

This work represents the first atmospheric characterization of an exoplanet using the high-resolution mode of the new Gemini High-resolution Optical SpecTrograph (GHOST) at the Gemini South Observatory. Our results, comparable to a previous analysis with HARPS \citep{Yan20}, highlight the efficacy of this instrument in detecting and characterizing exoplanet atmospheres at high spectral resolution. Although the observations presented herein were obtained during the instrument's System Verification run, GHOST is now fully commissioned and is available for general community use (Kalari et al. 2024, submitted). The addition of GHOST to Gemini's current suite of instruments opens the doors for numerous Southern hemisphere exoplanets which are particularly amenable to atmospheric characterization at high spectral resolution in the optical.

\section*{}
We thank Kim Venn, David Henderson, Pablo Prado, Christian Urrutia, Zachary Hartman, Janice Lee, Bryan Miller, Gabriel Perez, and the Gemini Observatory staff for their contributions to the GHOST System Verification. We also thank Joy Chavez and Karleyne Silva, the observer and operator on duty, for their support in obtaining these observations. We thank John Blakeslee for the helpful discussions. {Finally, we thank the anonymous referee for the thoughtful and constructive comments which helped improved the quality of this manuscript.}

Based on observations obtained under Program ID GS-2023A-SV-102, at the International Gemini Observatory, a program of NSF's NOIRLab, which is managed by the Association of Universities for Research in Astronomy (AURA) under a cooperative agreement with the National Science Foundation on behalf of the Gemini Observatory partnership: the National Science Foundation (United States), National Research Council (Canada), Agencia Nacional de Investigaci\'{o}n y Desarrollo (Chile), Ministerio de Ciencia, Tecnología e Innovaci\'{o}n (Argentina), Minist\'{e}rio da Ciência, Tecnologia, Inovações e Comunicações (Brazil), and Korea Astronomy and Space Science Institute (Republic of Korea). Data processed using DRAGONS (Data Reduction for Astronomy from Gemini Observatory North and
South).

GHOST was built by a collaboration between Australian Astronomical Optics at Macquarie University, National Research Council Herzberg of Canada, and the Australian National University, and funded by the International Gemini partnership. The instrument scientist is Dr.~Alan McConnachie at NRC, and the instrument team is also led by Dr.~Gordon Robertson (at AAO), and Dr.~Michael Ireland (at ANU). 

The authors would like to acknowledge the contributions of the GHOST instrument build team, the Gemini GHOST instrument team, the full SV team, and the rest of the Gemini operations team that were involved in making the SV observations a success.

E.K.D acknowledges the support of the Natural Sciences and Engineering Research Council of Canada (NSERC), funding reference number 568281-2022. J.D.T was supported for this work by NASA through the NASA Hubble Fellowship grant $\#$HST-HF2-51495.001-A awarded by the Space Telescope Science Institute, which is operated by the Association of Universities for Research in Astronomy, Incorporated, under NASA contract NAS5-26555. E.M. acknowledges funding from FAPEMIG under project number APQ-02493-22 and a research productivity grant number 309829/2022-4 awarded by the CNPq, Brazil.

This research has made use of the NASA Exoplanet Archive, which is operated by the California Institute of Technology, under contract with the National Aeronautics and Space Administration under the Exoplanet Exploration Program.


\vspace{5mm}
\facilities{Gemini:South (GHOST), Exoplanet Archive}

\software{\texttt{Astropy} \citep{astropy:2013, astropy:2018, astropy:2022}, \texttt{DRAGONS} \citep{dragons19,dragons22} \texttt{FastChem} \citep{Stock18}, \texttt{NumPy} \citep{numpy}, \texttt{petitRADTRANS} \citep{petitradtrans}, \texttt{SciPy} \citep{scipy}}

\appendix

\section{\textsc{SysRem} Results}
\label{app:sysrem}
The results of applying the \textsc{SysRem} algorithm to our data are displayed in Figs.~\ref{fig:sysrem_blue}, \ref{fig:sysrem_red_1}, and \ref{fig:sysrem_red_3}. Note that this does not include the orders which were excluded from our analysis, as described in Section \ref{subsec:obs}.

\begin{sidewaysfigure}
    \centering
    \includegraphics[width=\textwidth]{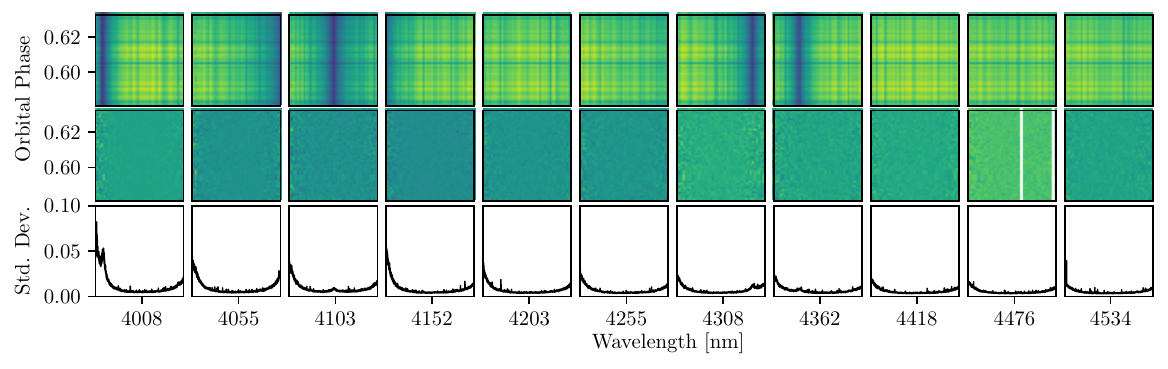}
        \includegraphics[width=\textwidth]{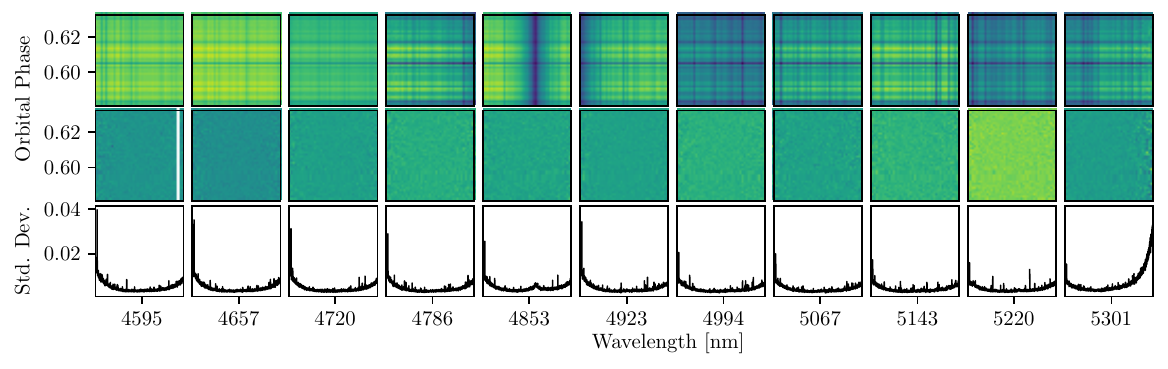}
        \caption{The results of applying the \textsc{SysRem} algorithm to the first 11 orders (top three rows) and last 11 orders (bottom three rows) in the blue arm of the spectrograph. {\textit{Top row:}} The data reduced by \texttt{DRAGONS}. Note that there may be some overlap in wavelength between subsequent orders; each order is displayed separately. {\textit{Middle row:}} The results of applying 8 iterations of \textsc{SysRem} to the data. At this point, the telluric and stellar absorption features have been removed, and the planetary signal is buried in the noise. {\textit{Bottom row:}} The standard deviation of the data after applying \textsc{SysRem}. The y-scale is the same for all plots in this row.}
    \label{fig:sysrem_blue}
\end{sidewaysfigure}

\begin{sidewaysfigure}
    \centering
    \includegraphics[width=\textwidth]{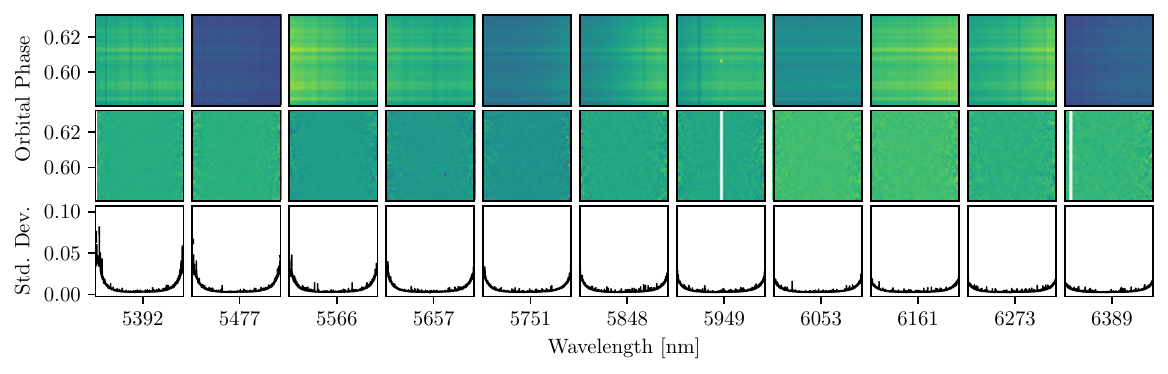}
        \includegraphics[width=\textwidth]{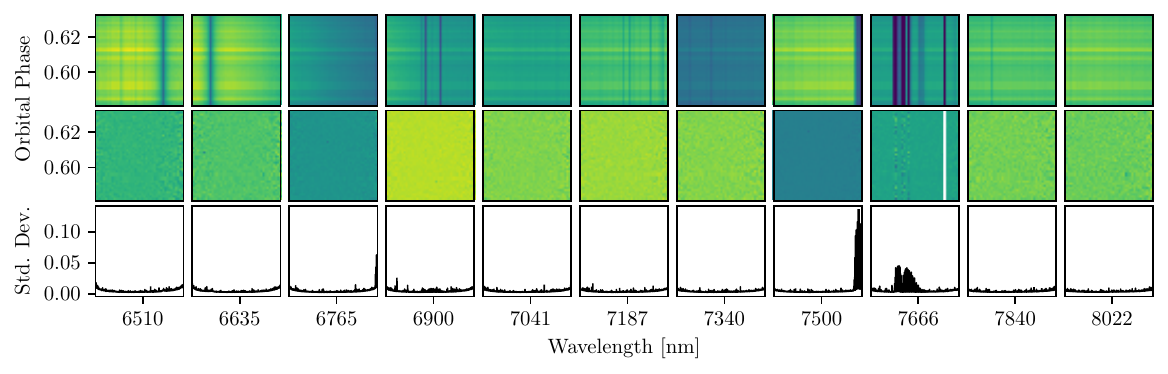}
    \caption{The same as Fig.~\ref{fig:sysrem_blue}, but for the first 11 orders (top three rows) and next 11 orders (bottom three rows) in the red arm of the spectrograph. As explained in Section \ref{subsec:tellurics}, we applied 6 iterations of \textsc{SysRem} to the red data.}
    \label{fig:sysrem_red_1}
\end{sidewaysfigure}

\begin{figure*}
    \centering
        \includegraphics[width=0.93\textwidth]{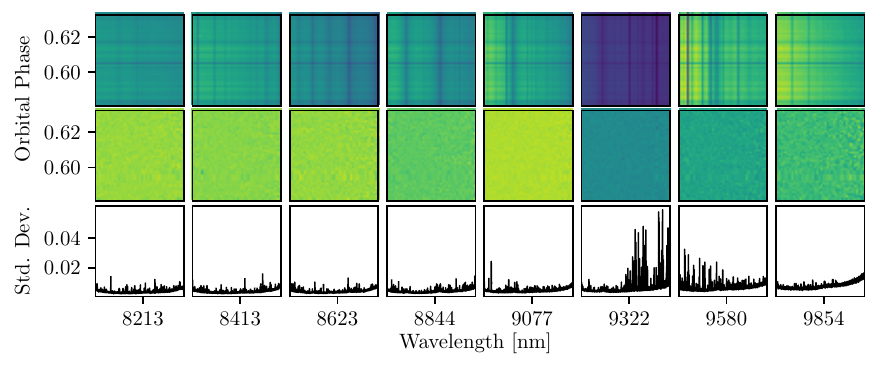}
    \caption{The same as Fig.~\ref{fig:sysrem_red_1}, but for the last 8 orders in the red.}
    \label{fig:sysrem_red_3}
\end{figure*}

\section{Model Line Lists}
\label{app:linelists}
{Table \ref{tab:linelists} presents the atmospheric species, line list reference, and whether the line list was initially in air or vacuum wavelengths, for all species used in our \texttt{petitRADTRANS} models. We note that all lists in vacuum wavelengths were converted to air wavelengths before our analysis in order to match the data.}

\begin{deluxetable*}{c c c}
\label{tab:linelists}
\tablecaption{Sources and wavelength frames for the line lists used to create the \texttt{petitRADTRANS} models in this work. Kurucz line lists: \url{http://kurucz.harvard.edu/}}
\tablehead{%
    \colhead{Atmospheric Species} & \colhead{Line List Reference} & \colhead{Air/Vacuum Wavelengths}
    }
\startdata
Fe & K.~Molaverdikhani/Kurucz
 & vacuum \\
Fe$^+$ & K.~Molaverdikhani/Kurucz & air \\
Ti & K.~Molaverdikhani/Kurucz & vacuum \\
V & K.~Molaverdikhani/Kurucz & vacuum \\
V$^+$ & K.~Molaverdikhani/Kurucz & vacuum \\
Al & K.~Molaverdikhani/Kurucz & vacuum \\
Ca & K.~Molaverdikhani/Kurucz & vacuum \\
Ca$^+$ & K.~Molaverdikhani/Kurucz & vacuum \\
Cr & K.~Molaverdikhani/Kurucz & vacuum \\
K & VALD with Allard wings (see \citealt{petitradtrans}) & vacuum \\
Mg & K.~Molaverdikhani/Kurucz & vacuum \\
Na & VALD with Allard wings (see \citealt{petitradtrans}) & vacuum \\
Si & K.~Molaverdikhani/Kurucz & vacuum \\
TiO & \cite{McKemmish19} & vacuum \\
VO & \cite{McKemmish16} & vacuum \\
FeH & Exomol (see \citealt{petitradtrans}) & vacuum \\
OH & HITEMP (see \citealt{petitradtrans}) & vacuum \\
\enddata
\end{deluxetable*}

\bibliography{references}{}

\begin{thebibliography}{}
\expandafter\ifx\csname natexlab\endcsname\relax\def\natexlab#1{#1}\fi
\providecommand{\url}[1]{\href{#1}{#1}}
\providecommand{\dodoi}[1]{doi:~\href{http://doi.org/#1}{\nolinkurl{#1}}}
\providecommand{\doeprint}[1]{\href{http://ascl.net/#1}{\nolinkurl{http://ascl.net/#1}}}
\providecommand{\doarXiv}[1]{\href{https://arxiv.org/abs/#1}{\nolinkurl{https://arxiv.org/abs/#1}}}

\bibitem[{{Anderson} {et~al.}(2018){Anderson}, {Temple}, {Nielsen}, {Burdanov}, {Hellier}, {Bouchy}, {Brown}, {Collier Cameron}, {Gillon}, {Jehin}, {Maxted}, {Pepe}, {Pollacco}, {Pozuelos}, {Queloz}, {S{\'e}gransan}, {Smalley}, {Triaud}, {Turner}, {Udry}, \& {West}}]{Anderson18}
{Anderson}, D.~R., {Temple}, L.~Y., {Nielsen}, L.~D., {et~al.} 2018, arXiv e-prints, arXiv:1809.04897, \dodoi{10.48550/arXiv.1809.04897}

\bibitem[{{Astropy Collaboration} {et~al.}(2013){Astropy Collaboration}, {Robitaille}, {Tollerud}, {Greenfield}, {Droettboom}, {Bray}, {Aldcroft}, {Davis}, {Ginsburg}, {Price-Whelan}, {Kerzendorf}, {Conley}, {Crighton}, {Barbary}, {Muna}, {Ferguson}, {Grollier}, {Parikh}, {Nair}, {Unther}, {Deil}, {Woillez}, {Conseil}, {Kramer}, {Turner}, {Singer}, {Fox}, {Weaver}, {Zabalza}, {Edwards}, {Azalee Bostroem}, {Burke}, {Casey}, {Crawford}, {Dencheva}, {Ely}, {Jenness}, {Labrie}, {Lim}, {Pierfederici}, {Pontzen}, {Ptak}, {Refsdal}, {Servillat}, \& {Streicher}}]{astropy:2013}
{Astropy Collaboration}, {Robitaille}, T.~P., {Tollerud}, E.~J., {et~al.} 2013, \aap, 558, A33, \dodoi{10.1051/0004-6361/201322068}

\bibitem[{{Astropy Collaboration} {et~al.}(2018){Astropy Collaboration}, {Price-Whelan}, {Sip{\H{o}}cz}, {G{\"u}nther}, {Lim}, {Crawford}, {Conseil}, {Shupe}, {Craig}, {Dencheva}, {Ginsburg}, {Vand erPlas}, {Bradley}, {P{\'e}rez-Su{\'a}rez}, {de Val-Borro}, {Aldcroft}, {Cruz}, {Robitaille}, {Tollerud}, {Ardelean}, {Babej}, {Bach}, {Bachetti}, {Bakanov}, {Bamford}, {Barentsen}, {Barmby}, {Baumbach}, {Berry}, {Biscani}, {Boquien}, {Bostroem}, {Bouma}, {Brammer}, {Bray}, {Breytenbach}, {Buddelmeijer}, {Burke}, {Calderone}, {Cano Rodr{\'\i}guez}, {Cara}, {Cardoso}, {Cheedella}, {Copin}, {Corrales}, {Crichton}, {D'Avella}, {Deil}, {Depagne}, {Dietrich}, {Donath}, {Droettboom}, {Earl}, {Erben}, {Fabbro}, {Ferreira}, {Finethy}, {Fox}, {Garrison}, {Gibbons}, {Goldstein}, {Gommers}, {Greco}, {Greenfield}, {Groener}, {Grollier}, {Hagen}, {Hirst}, {Homeier}, {Horton}, {Hosseinzadeh}, {Hu}, {Hunkeler}, {Ivezi{\'c}}, {Jain}, {Jenness}, {Kanarek}, {Kendrew}, {Kern}, {Kerzendorf}, {Khvalko}, {King}, {Kirkby}, {Kulkarni}, {Kumar}, {Lee}, {Lenz}, {Littlefair}, {Ma}, {Macleod}, {Mastropietro}, {McCully}, {Montagnac}, {Morris}, {Mueller}, {Mumford}, {Muna}, {Murphy}, {Nelson}, {Nguyen}, {Ninan}, {N{\"o}the}, {Ogaz}, {Oh}, {Parejko}, {Parley}, {Pascual}, {Patil}, {Patil}, {Plunkett}, {Prochaska}, {Rastogi}, {Reddy Janga}, {Sabater}, {Sakurikar}, {Seifert}, {Sherbert}, {Sherwood-Taylor}, {Shih}, {Sick}, {Silbiger}, {Singanamalla}, {Singer}, {Sladen}, {Sooley}, {Sornarajah}, {Streicher}, {Teuben}, {Thomas}, {Tremblay}, {Turner}, {Terr{\'o}n}, {van Kerkwijk}, {de la Vega}, {Watkins}, {Weaver}, {Whitmore}, {Woillez}, {Zabalza}, \& {Astropy Contributors}}]{astropy:2018}
{Astropy Collaboration}, {Price-Whelan}, A.~M., {Sip{\H{o}}cz}, B.~M., {et~al.} 2018, \aj, 156, 123, \dodoi{10.3847/1538-3881/aabc4f}

\bibitem[{{Astropy Collaboration} {et~al.}(2022){Astropy Collaboration}, {Price-Whelan}, {Lim}, {Earl}, {Starkman}, {Bradley}, {Shupe}, {Patil}, {Corrales}, {Brasseur}, {N{"o}the}, {Donath}, {Tollerud}, {Morris}, {Ginsburg}, {Vaher}, {Weaver}, {Tocknell}, {Jamieson}, {van Kerkwijk}, {Robitaille}, {Merry}, {Bachetti}, {G{"u}nther}, {Aldcroft}, {Alvarado-Montes}, {Archibald}, {B{'o}di}, {Bapat}, {Barentsen}, {Baz{'a}n}, {Biswas}, {Boquien}, {Burke}, {Cara}, {Cara}, {Conroy}, {Conseil}, {Craig}, {Cross}, {Cruz}, {D'Eugenio}, {Dencheva}, {Devillepoix}, {Dietrich}, {Eigenbrot}, {Erben}, {Ferreira}, {Foreman-Mackey}, {Fox}, {Freij}, {Garg}, {Geda}, {Glattly}, {Gondhalekar}, {Gordon}, {Grant}, {Greenfield}, {Groener}, {Guest}, {Gurovich}, {Handberg}, {Hart}, {Hatfield-Dodds}, {Homeier}, {Hosseinzadeh}, {Jenness}, {Jones}, {Joseph}, {Kalmbach}, {Karamehmetoglu}, {Ka{l}uszy{'n}ski}, {Kelley}, {Kern}, {Kerzendorf}, {Koch}, {Kulumani}, {Lee}, {Ly}, {Ma}, {MacBride}, {Maljaars}, {Muna}, {Murphy}, {Norman}, {O'Steen}, {Oman}, {Pacifici}, {Pascual}, {Pascual-Granado}, {Patil}, {Perren}, {Pickering}, {Rastogi}, {Roulston}, {Ryan}, {Rykoff}, {Sabater}, {Sakurikar}, {Salgado}, {Sanghi}, {Saunders}, {Savchenko}, {Schwardt}, {Seifert-Eckert}, {Shih}, {Jain}, {Shukla}, {Sick}, {Simpson}, {Singanamalla}, {Singer}, {Singhal}, {Sinha}, {Sip{H{o}}cz}, {Spitler}, {Stansby}, {Streicher}, {{{S}}umak}, {Swinbank}, {Taranu}, {Tewary}, {Tremblay}, {Val-Borro}, {Van Kooten}, {Vasovi{'c}}, {Verma}, {de Miranda Cardoso}, {Williams}, {Wilson}, {Winkel}, {Wood-Vasey}, {Xue}, {Yoachim}, {Zhang}, {Zonca}, \& {Astropy Project Contributors}}]{astropy:2022}
{Astropy Collaboration}, {Price-Whelan}, A.~M., {Lim}, P.~L., {et~al.} 2022, \apj, 935, 167, \dodoi{10.3847/1538-4357/ac7c74}

\bibitem[{{Bell} \& {Cowan}(2018)}]{Bell18}
{Bell}, T.~J., \& {Cowan}, N.~B. 2018, \apjl, 857, L20, \dodoi{10.3847/2041-8213/aabcc8}

\bibitem[{{Bello-Arufe} {et~al.}(2022){Bello-Arufe}, {Buchhave}, {Mendon{\c{c}}a}, {Tronsgaard}, {Heng}, {Jens Hoeijmakers}, \& {Mayo}}]{BelloArufe22}
{Bello-Arufe}, A., {Buchhave}, L.~A., {Mendon{\c{c}}a}, J.~M., {et~al.} 2022, \aap, 662, A51, \dodoi{10.1051/0004-6361/202142787}

\bibitem[{{Birkby} {et~al.}(2013){Birkby}, {de Kok}, {Brogi}, {de Mooij}, {Schwarz}, {Albrecht}, \& {Snellen}}]{Birkby13}
{Birkby}, J.~L., {de Kok}, R.~J., {Brogi}, M., {et~al.} 2013, \mnras, 436, L35, \dodoi{10.1093/mnrasl/slt107}

\bibitem[{{Boldt-Christmas} {et~al.}(2023){Boldt-Christmas}, {Lesjak}, {Wehrhahn}, {Piskunov}, {Rains}, {Nortmann}, \& {Kochukhov}}]{BoldtChristmas23}
{Boldt-Christmas}, L., {Lesjak}, F., {Wehrhahn}, A., {et~al.} 2023, arXiv e-prints, arXiv:2312.08320, \dodoi{10.48550/arXiv.2312.08320}

\bibitem[{{Brogi} {et~al.}(2014){Brogi}, {de Kok}, {Birkby}, {Schwarz}, \& {Snellen}}]{Brogi14}
{Brogi}, M., {de Kok}, R.~J., {Birkby}, J.~L., {Schwarz}, H., \& {Snellen}, I.~A.~G. 2014, \aap, 565, A124, \dodoi{10.1051/0004-6361/201423537}

\bibitem[{{Brogi} \& {Line}(2019)}]{Brogi19}
{Brogi}, M., \& {Line}, M.~R. 2019, \aj, 157, 114, \dodoi{10.3847/1538-3881/aaffd3}

\bibitem[{{Brogi} {et~al.}(2013){Brogi}, {Snellen}, {de Kok}, {Albrecht}, {Birkby}, \& {de Mooij}}]{Brogi13}
{Brogi}, M., {Snellen}, I.~A.~G., {de Kok}, R.~J., {et~al.} 2013, \apj, 767, 27, \dodoi{10.1088/0004-637X/767/1/27}

\bibitem[{{Cabot} {et~al.}(2019){Cabot}, {Madhusudhan}, {Hawker}, \& {Gandhi}}]{Cabot19}
{Cabot}, S. H.~C., {Madhusudhan}, N., {Hawker}, G.~A., \& {Gandhi}, S. 2019, \mnras, 482, 4422, \dodoi{10.1093/mnras/sty2994}

\bibitem[{{Cauley} {et~al.}(2020){Cauley}, {Shkolnik}, {Ilyin}, {Strassmeier}, {Redfield}, \& {Jensen}}]{Cauley20}
{Cauley}, P.~W., {Shkolnik}, E.~L., {Ilyin}, I., {et~al.} 2020, Research Notes of the American Astronomical Society, 4, 53, \dodoi{10.3847/2515-5172/ab88dc}

\bibitem[{{Chatzikos} {et~al.}(2023){Chatzikos}, {Bianchi}, {Camilloni}, {Chakraborty}, {Gunasekera}, {Guzm{\'a}n}, {Milby}, {Sarkar}, {Shaw}, {van Hoof}, \& {Ferland}}]{cloudy}
{Chatzikos}, M., {Bianchi}, S., {Camilloni}, F., {et~al.} 2023, \rmxaa, 59, 327, \dodoi{10.22201/ia.01851101p.2023.59.02.12}

\bibitem[{{Cheverall} {et~al.}(2023){Cheverall}, {Madhusudhan}, \& {Holmberg}}]{Cheverall23}
{Cheverall}, C.~J., {Madhusudhan}, N., \& {Holmberg}, M. 2023, \mnras, 522, 661, \dodoi{10.1093/mnras/stad648}

\bibitem[{{Cont} {et~al.}(2021){Cont}, {Yan}, {Reiners}, {Casasayas-Barris}, {Molli{\`e}re}, {Pall{\'e}}, {Henning}, {Nortmann}, {Stangret}, {Czesla}, {L{\'o}pez-Puertas}, {S{\'a}nchez-L{\'o}pez}, {Rodler}, {Ribas}, {Quirrenbach}, {Caballero}, {Amado}, {Carone}, {Khaimova}, {Kreidberg}, {Molaverdikhani}, {Montes}, {Morello}, {Nagel}, {Oshagh}, \& {Zechmeister}}]{Cont21}
{Cont}, D., {Yan}, F., {Reiners}, A., {et~al.} 2021, \aap, 651, A33, \dodoi{10.1051/0004-6361/202140732}

\bibitem[{{Deibert} {et~al.}(2021){Deibert}, {de Mooij}, {Jayawardhana}, {Turner}, {Ridden-Harper}, {Fossati}, {Hood}, {Fortney}, {Flagg}, {MacDonald}, {Allart}, \& {Sing}}]{Deibert21}
{Deibert}, E.~K., {de Mooij}, E. J.~W., {Jayawardhana}, R., {et~al.} 2021, \apjl, 919, L15, \dodoi{10.3847/2041-8213/ac2513}

\bibitem[{{Deibert} {et~al.}(2023){Deibert}, {de Mooij}, {Jayawardhana}, {Turner}, {Ridden-Harper}, {Hood}, {Fortney}, {Flagg}, {Fossati}, {Allart}, {Brogi}, \& {MacDonald}}]{Deibert23}
---. 2023, \aj, 166, 141, \dodoi{10.3847/1538-3881/acebdc}

\bibitem[{{Deline} {et~al.}(2022){Deline}, {Hooton}, {Lendl}, {Morris}, {Salmon}, {Olofsson}, {Broeg}, {Ehrenreich}, {Beck}, {Brandeker}, {Hoyer}, {Sulis}, {Van Grootel}, {Bourrier}, {Demangeon}, {Demory}, {Heng}, {Parviainen}, {Serrano}, {Singh}, {Bonfanti}, {Fossati}, {Kitzmann}, {Sousa}, {Wilson}, {Alibert}, {Alonso}, {Anglada}, {B{\'a}rczy}, {Barrado Navascues}, {Barros}, {Baumjohann}, {Beck}, {Bekkelien}, {Benz}, {Billot}, {Bonfils}, {Cabrera}, {Charnoz}, {Collier Cameron}, {Corral van Damme}, {Csizmadia}, {Davies}, {Deleuil}, {Delrez}, {de Roche}, {Erikson}, {Fortier}, {Fridlund}, {Futyan}, {Gandolfi}, {Gillon}, {G{\"u}del}, {Gutermann}, {Hasiba}, {Isaak}, {Kiss}, {Laskar}, {Lecavelier des Etangs}, {Lovis}, {Magrin}, {Maxted}, {Munari}, {Nascimbeni}, {Ottensamer}, {Pagano}, {Pall{\'e}}, {Peter}, {Piotto}, {Pollacco}, {Queloz}, {Ragazzoni}, {Rando}, {Rauer}, {Ribas}, {Santos}, {Scandariato}, {S{\'e}gransan}, {Simon}, {Smith}, {Steller}, {Szab{\'o}}, {Thomas}, {Udry}, {Walter}, \& {Walton}}]{Deline22}
{Deline}, A., {Hooton}, M.~J., {Lendl}, M., {et~al.} 2022, \aap, 659, A74, \dodoi{10.1051/0004-6361/202142400}

\bibitem[{{Ehrenreich} {et~al.}(2020){Ehrenreich}, {Lovis}, {Allart}, {Zapatero Osorio}, {Pepe}, {Cristiani}, {Rebolo}, {Santos}, {Borsa}, {Demangeon}, {Dumusque}, {Gonz{\'a}lez Hern{\'a}ndez}, {Casasayas-Barris}, {S{\'e}gransan}, {Sousa}, {Abreu}, {Adibekyan}, {Affolter}, {Allende Prieto}, {Alibert}, {Aliverti}, {Alves}, {Amate}, {Avila}, {Baldini}, {Bandy}, {Benz}, {Bianco}, {Bolmont}, {Bouchy}, {Bourrier}, {Broeg}, {Cabral}, {Calderone}, {Pall{\'e}}, {Cegla}, {Cirami}, {Coelho}, {Conconi}, {Coretti}, {Cumani}, {Cupani}, {Dekker}, {Delabre}, {Deiries}, {D'Odorico}, {Di Marcantonio}, {Figueira}, {Fragoso}, {Genolet}, {Genoni}, {G{\'e}nova Santos}, {Hara}, {Hughes}, {Iwert}, {Kerber}, {Knudstrup}, {Landoni}, {Lavie}, {Lizon}, {Lendl}, {Lo Curto}, {Maire}, {Manescau}, {Martins}, {M{\'e}gevand}, {Mehner}, {Micela}, {Modigliani}, {Molaro}, {Monteiro}, {Monteiro}, {Moschetti}, {M{\"u}ller}, {Nunes}, {Oggioni}, {Oliveira}, {Pariani}, {Pasquini}, {Poretti}, {Rasilla}, {Redaelli}, {Riva}, {Santana Tschudi}, {Santin}, {Santos}, {Segovia Milla}, {Seidel}, {Sosnowska}, {Sozzetti}, {Span{\`o}}, {Su{\'a}rez Mascare{\~n}o}, {Tabernero}, {Tenegi}, {Udry}, {Zanutta}, \& {Zerbi}}]{Ehrenreich20}
{Ehrenreich}, D., {Lovis}, C., {Allart}, R., {et~al.} 2020, \nat, 580, 597, \dodoi{10.1038/s41586-020-2107-1}

\bibitem[{{Esteves} {et~al.}(2017){Esteves}, {de Mooij}, {Jayawardhana}, {Watson}, \& {de Kok}}]{Esteves17}
{Esteves}, L.~J., {de Mooij}, E. J.~W., {Jayawardhana}, R., {Watson}, C., \& {de Kok}, R. 2017, \aj, 153, 268, \dodoi{10.3847/1538-3881/aa7133}

\bibitem[{{Gandhi} {et~al.}(2023){Gandhi}, {Kesseli}, {Zhang}, {Louca}, {Snellen}, {Brogi}, {Miguel}, {Casasayas-Barris}, {Pelletier}, {Landman}, {Maguire}, \& {Gibson}}]{Gandhi23}
{Gandhi}, S., {Kesseli}, A., {Zhang}, Y., {et~al.} 2023, \aj, 165, 242, \dodoi{10.3847/1538-3881/accd65}

\bibitem[{{Gibson} {et~al.}(2022){Gibson}, {Nugroho}, {Lothringer}, {Maguire}, \& {Sing}}]{Gibson22}
{Gibson}, N.~P., {Nugroho}, S.~K., {Lothringer}, J., {Maguire}, C., \& {Sing}, D.~K. 2022, \mnras, 512, 4618, \dodoi{10.1093/mnras/stac091}

\bibitem[{Harris {et~al.}(2020)Harris, Millman, van~der Walt, Gommers, Virtanen, Cournapeau, Wieser, Taylor, Berg, Smith, Kern, Picus, Hoyer, van Kerkwijk, Brett, Haldane, del R{\'{i}}o, Wiebe, Peterson, G{\'{e}}rard-Marchant, Sheppard, Reddy, Weckesser, Abbasi, Gohlke, \& Oliphant}]{numpy}
Harris, C.~R., Millman, K.~J., van~der Walt, S.~J., {et~al.} 2020, Nature, 585, 357, \dodoi{10.1038/s41586-020-2649-2}

\bibitem[{{Hayes} {et~al.}(2022){Hayes}, {Waller}, {Ireland}, {Nielsen}, {White}, {Bento}, {Venn}, {Pazder}, {McConnachie}, {Simpson}, \& {Labrie}}]{Hayes22}
{Hayes}, C.~R., {Waller}, F., {Ireland}, M., {et~al.} 2022, in Society of Photo-Optical Instrumentation Engineers (SPIE) Conference Series, Vol. 12184, Ground-based and Airborne Instrumentation for Astronomy IX, ed. C.~J. {Evans}, J.~J. {Bryant}, \& K.~{Motohara}, 121846H, \dodoi{10.1117/12.2642905}

\bibitem[{{Herman} {et~al.}(2020){Herman}, {de Mooij}, {Jayawardhana}, \& {Brogi}}]{Herman20}
{Herman}, M.~K., {de Mooij}, E. J.~W., {Jayawardhana}, R., \& {Brogi}, M. 2020, \aj, 160, 93, \dodoi{10.3847/1538-3881/ab9e77}

\bibitem[{{Herman} {et~al.}(2022){Herman}, {de Mooij}, {Nugroho}, {Gibson}, \& {Jayawardhana}}]{Herman22}
{Herman}, M.~K., {de Mooij}, E. J.~W., {Nugroho}, S.~K., {Gibson}, N.~P., \& {Jayawardhana}, R. 2022, \aj, 163, 248, \dodoi{10.3847/1538-3881/ac5f4d}

\bibitem[{{Hoeijmakers} {et~al.}(2022){Hoeijmakers}, {Kitzmann}, {Morris}, {Prinoth}, {Borsato}, {Thorsbro}, {Pino}, {Lee}, {Ak{\i}n}, {Seidel}, {Birkby}, {Allart}, \& {Heng}}]{Hoeijmakers22}
{Hoeijmakers}, H.~J., {Kitzmann}, D., {Morris}, B.~M., {et~al.} 2022, arXiv e-prints, arXiv:2210.12847, \dodoi{10.48550/arXiv.2210.12847}

\bibitem[{{Ireland} {et~al.}(2014){Ireland}, {Anthony}, {Burley}, {Chisholm}, {Churilov}, {Dunn}, {Frost}, {Lawrence}, {Loop}, {McGregor}, {Martell}, {McConnachie}, {McDermid}, {Pazder}, {Reshetov}, {Robertson}, {Sheinis}, {Tims}, {Young}, \& {Zhelem}}]{Ireland14}
{Ireland}, M., {Anthony}, A., {Burley}, G., {et~al.} 2014, in Society of Photo-Optical Instrumentation Engineers (SPIE) Conference Series, Vol. 9147, Ground-based and Airborne Instrumentation for Astronomy V, ed. S.~K. {Ramsay}, I.~S. {McLean}, \& H.~{Takami}, 91471J, \dodoi{10.1117/12.2057356}

\bibitem[{{Ireland} {et~al.}(2018){Ireland}, {White}, {Bento}, {Farrell}, {Labrie}, {Luvaul}, {Nielsen}, \& {Simpson}}]{Ireland18}
{Ireland}, M.~J., {White}, M., {Bento}, J.~P., {et~al.} 2018, in Society of Photo-Optical Instrumentation Engineers (SPIE) Conference Series, Vol. 10707, Software and Cyberinfrastructure for Astronomy V, ed. J.~C. {Guzman} \& J.~{Ibsen}, 1070735, \dodoi{10.1117/12.2314418}

\bibitem[{{Ireland} {et~al.}(2012){Ireland}, {Barnes}, {Cochrane}, {Colless}, {Connor}, {Horton}, {Gibson}, {Lawrence}, {Martell}, {McGregor}, {Nicolle}, {Nield}, {Orr}, {Robertson}, {Ryder}, {Sheinis}, {Smith}, {Staszak}, {Tims}, {Xavier}, {Young}, \& {Zheng}}]{Ireland12}
{Ireland}, M.~J., {Barnes}, S., {Cochrane}, D., {et~al.} 2012, in Society of Photo-Optical Instrumentation Engineers (SPIE) Conference Series, Vol. 8446, Ground-based and Airborne Instrumentation for Astronomy IV, ed. I.~S. {McLean}, S.~K. {Ramsay}, \& H.~{Takami}, 844629, \dodoi{10.1117/12.925746}

\bibitem[{{Ivshina} \& {Winn}(2022)}]{Ivshina22}
{Ivshina}, E.~S., \& {Winn}, J.~N. 2022, \apjs, 259, 62, \dodoi{10.3847/1538-4365/ac545b}

\bibitem[{{Johnson} {et~al.}(2023){Johnson}, {Wang}, {Asnodkar}, {Bonomo}, {Gaudi}, {Henning}, {Ilyin}, {Keles}, {Malavolta}, {Mallonn}, {Molaverdikhani}, {Nascimbeni}, {Patience}, {Poppenhaeger}, {Scandariato}, {Schlawin}, {Shkolnik}, {Sicilia}, {Sozzetti}, {Strassmeier}, {Veillet}, \& {Yan}}]{Johnson23}
{Johnson}, M.~C., {Wang}, J., {Asnodkar}, A.~P., {et~al.} 2023, \aj, 165, 157, \dodoi{10.3847/1538-3881/acb7e2}

\bibitem[{{Kasper} {et~al.}(2021){Kasper}, {Bean}, {Line}, {Seifahrt}, {St{\"u}rmer}, {Pino}, {D{\'e}sert}, \& {Brogi}}]{Kasper21}
{Kasper}, D., {Bean}, J.~L., {Line}, M.~R., {et~al.} 2021, \apjl, 921, L18, \dodoi{10.3847/2041-8213/ac30e1}

\bibitem[{{Komacek} \& {Tan}(2018)}]{Komacek18}
{Komacek}, T.~D., \& {Tan}, X. 2018, Research Notes of the American Astronomical Society, 2, 36, \dodoi{10.3847/2515-5172/aac5e7}

\bibitem[{{Labrie} {et~al.}(2019){Labrie}, {Anderson}, {C{\'a}rdenes}, {Simpson}, \& {Turner}}]{dragons19}
{Labrie}, K., {Anderson}, K., {C{\'a}rdenes}, R., {Simpson}, C., \& {Turner}, J. E.~H. 2019, in Astronomical Society of the Pacific Conference Series, Vol. 523, Astronomical Data Analysis Software and Systems XXVII, ed. P.~J. {Teuben}, M.~W. {Pound}, B.~A. {Thomas}, \& E.~M. {Warner}, 321

\bibitem[{{Labrie} {et~al.}(2022){Labrie}, {Simpson}, {Anderson}, {Cardenas}, {Turner}, {Quint}, {Conseil}, \& {Oberdorf}}]{dragons22}
{Labrie}, K., {Simpson}, C., {Anderson}, K., {et~al.} 2022, {DRAGONS}, 3.0.4, Zenodo,  Zenodo, \dodoi{10.5281/zenodo.7308726}

\bibitem[{{Langeveld} {et~al.}(2022){Langeveld}, {Madhusudhan}, \& {Cabot}}]{Langeveld22}
{Langeveld}, A.~B., {Madhusudhan}, N., \& {Cabot}, S. H.~C. 2022, \mnras, 514, 5192, \dodoi{10.1093/mnras/stac1539}

\bibitem[{{Lee} {et~al.}(2022){Lee}, {Prinoth}, {Kitzmann}, {Tsai}, {Hoeijmakers}, {Borsato}, \& {Heng}}]{Lee22}
{Lee}, E. K.~H., {Prinoth}, B., {Kitzmann}, D., {et~al.} 2022, \mnras, 517, 240, \dodoi{10.1093/mnras/stac2246}

\bibitem[{{Lendl} {et~al.}(2020){Lendl}, {Csizmadia}, {Deline}, {Fossati}, {Kitzmann}, {Heng}, {Hoyer}, {Salmon}, {Benz}, {Broeg}, {Ehrenreich}, {Fortier}, {Queloz}, {Bonfanti}, {Brandeker}, {Collier Cameron}, {Delrez}, {Garcia Mu{\~n}oz}, {Hooton}, {Maxted}, {Morris}, {Van Grootel}, {Wilson}, {Alibert}, {Alonso}, {Asquier}, {Bandy}, {B{\'a}rczy}, {Barrado}, {Barros}, {Baumjohann}, {Beck}, {Beck}, {Bekkelien}, {Bergomi}, {Billot}, {Biondi}, {Bonfils}, {Bourrier}, {Busch}, {Cabrera}, {Cessa}, {Charnoz}, {Chazelas}, {Corral Van Damme}, {Davies}, {Deleuil}, {Demangeon}, {Demory}, {Erikson}, {Farinato}, {Fridlund}, {Futyan}, {Gandolfi}, {Gillon}, {Guterman}, {Hasiba}, {Hernandez}, {Isaak}, {Kiss}, {Kuntzer}, {Lecavelier des Etangs}, {L{\"u}ftinger}, {Laskar}, {Lovis}, {Magrin}, {Malvasio}, {Marafatto}, {Michaelis}, {Munari}, {Nascimbeni}, {Olofsson}, {Ottacher}, {Ottensamer}, {Pagano}, {Pall{\'e}}, {Peter}, {Piazza}, {Piotto}, {Pollacco}, {Ratti}, {Rauer}, {Ragazzoni}, {Rando}, {Ribas}, {Rieder}, {Rohlfs}, {Safa}, {Santos}, {Scandariato}, {S{\'e}gransan}, {Simon}, {Singh}, {Smith}, {Sordet}, {Sousa}, {Steller}, {Szab{\'o}}, {Thomas}, {Tschentscher}, {Udry}, {Viotto}, {Walter}, {Walton}, {Wildi}, \& {Wolter}}]{Lendl20}
{Lendl}, M., {Csizmadia}, S., {Deline}, A., {et~al.} 2020, \aap, 643, A94, \dodoi{10.1051/0004-6361/202038677}

\bibitem[{{Lothringer} \& {Barman}(2019)}]{Lothringer19}
{Lothringer}, J.~D., \& {Barman}, T. 2019, \apj, 876, 69, \dodoi{10.3847/1538-4357/ab1485}

\bibitem[{{Lothringer} {et~al.}(2018){Lothringer}, {Barman}, \& {Koskinen}}]{Lothringer18}
{Lothringer}, J.~D., {Barman}, T., \& {Koskinen}, T. 2018, \apj, 866, 27, \dodoi{10.3847/1538-4357/aadd9e}

\bibitem[{{McConnachie} {et~al.}(2022{\natexlab{a}}){McConnachie}, {Hayes}, {Pazder}, {Ireland}, {Kalari}, {McDermid}, \& {Margheim}}]{GHOST22}
{McConnachie}, A.~W., {Hayes}, C.~R., {Pazder}, J., {et~al.} 2022{\natexlab{a}}, Nature Astronomy, 6, 1491, \dodoi{10.1038/s41550-022-01854-1}

\bibitem[{{McConnachie} {et~al.}(2022{\natexlab{b}}){McConnachie}, {Hayes}, {Ireland}, {Waller}, {Berg}, {Pazder}, {Margheim}, {Kalari}, {Farrell}, \& {Robertson}}]{McConnachie22}
{McConnachie}, A.~W., {Hayes}, C., {Ireland}, M., {et~al.} 2022{\natexlab{b}}, in Society of Photo-Optical Instrumentation Engineers (SPIE) Conference Series, Vol. 12184, Ground-based and Airborne Instrumentation for Astronomy IX, ed. C.~J. {Evans}, J.~J. {Bryant}, \& K.~{Motohara}, 121841E, \dodoi{10.1117/12.2630407}

\bibitem[{{McConnachie} {et~al.}(2024){McConnachie}, {Hayes}, {Robertson}, {Pazder}, {Ireland}, {Burley}, {Churilov}, {Lothrop}, {Zhelem}, {Kalari}, {Anthony}, {Baker}, {Berg}, {Chapin}, {Chin}, {Densmore}, {Diaz}, {Dunn}, {Edgar}, {Farrell}, {Firpo}, {Fuentes}, {Gomez-Jimenez}, {Hardy}, {Henderson}, {Hill}, {Labrie}, {Jensen}, {Lambert}, {Lawrence}, {Macdonald}, {Margheim}, {Millar}, {Muller}, {Nielsen}, {P{\'e}rez}, {Quiroz}, {Ruiz-Carmona}, {Sebo}, {Sestito}, {Silva}, {Simpson}, {Smith}, {Venkatesan}, {Waller}, {Waller}, {Wevers}, {Venn}, {Young}, \& {Silversides}}]{McConnachie24}
{McConnachie}, A.~W., {Hayes}, C.~R., {Robertson}, J.~G., {et~al.} 2024, \pasp, 136, 035001, \dodoi{10.1088/1538-3873/ad1ed4}

\bibitem[{{McKemmish} {et~al.}(2019){McKemmish}, {Masseron}, {Hoeijmakers}, {P{\'e}rez-Mesa}, {Grimm}, {Yurchenko}, \& {Tennyson}}]{McKemmish19}
{McKemmish}, L.~K., {Masseron}, T., {Hoeijmakers}, H.~J., {et~al.} 2019, \mnras, 488, 2836, \dodoi{10.1093/mnras/stz1818}

\bibitem[{{McKemmish} {et~al.}(2016){McKemmish}, {Yurchenko}, \& {Tennyson}}]{McKemmish16}
{McKemmish}, L.~K., {Yurchenko}, S.~N., \& {Tennyson}, J. 2016, \mnras, 463, 771, \dodoi{10.1093/mnras/stw1969}

\bibitem[{{Molli{\`e}re} {et~al.}(2019){Molli{\`e}re}, {Wardenier}, {van Boekel}, {Henning}, {Molaverdikhani}, \& {Snellen}}]{petitradtrans}
{Molli{\`e}re}, P., {Wardenier}, J.~P., {van Boekel}, R., {et~al.} 2019, \aap, 627, A67, \dodoi{10.1051/0004-6361/201935470}

\bibitem[{{Nugroho} {et~al.}(2020){Nugroho}, {Gibson}, {de Mooij}, {Herman}, {Watson}, {Kawahara}, \& {Merritt}}]{Nugroho20}
{Nugroho}, S.~K., {Gibson}, N.~P., {de Mooij}, E. J.~W., {et~al.} 2020, \apjl, 898, L31, \dodoi{10.3847/2041-8213/aba4b6}

\bibitem[{{Parmentier} {et~al.}(2018){Parmentier}, {Line}, {Bean}, {Mansfield}, {Kreidberg}, {Lupu}, {Visscher}, {D{\'e}sert}, {Fortney}, {Deleuil}, {Arcangeli}, {Showman}, \& {Marley}}]{Parmentier18}
{Parmentier}, V., {Line}, M.~R., {Bean}, J.~L., {et~al.} 2018, \aap, 617, A110, \dodoi{10.1051/0004-6361/201833059}

\bibitem[{{Pelletier} {et~al.}(2023){Pelletier}, {Benneke}, {Ali-Dib}, {Prinoth}, {Kasper}, {Seifahrt}, {Bean}, {Debras}, {Klein}, {Bazinet}, {Hoeijmakers}, {Kesseli}, {Lim}, {Carmona}, {Pino}, {Casasayas-Barris}, {Hood}, \& {St{\"u}rmer}}]{Pelletier23}
{Pelletier}, S., {Benneke}, B., {Ali-Dib}, M., {et~al.} 2023, \nat, 619, 491, \dodoi{10.1038/s41586-023-06134-0}

\bibitem[{{Petz} {et~al.}(2023){Petz}, {Johnson}, {Pai Asnodkar}, {Wang}, {Gaudi}, {Henning}, {Keles}, {Molaverdikhani}, {Poppenhaeger}, {Scandariato}, {Shkolnik}, {Sicilia}, {Strassmeier}, \& {Yan}}]{Petz23}
{Petz}, S., {Johnson}, M.~C., {Pai Asnodkar}, A., {et~al.} 2023, arXiv e-prints, arXiv:2310.09352, \dodoi{10.48550/arXiv.2310.09352}

\bibitem[{{Pino} {et~al.}(2020){Pino}, {D{\'e}sert}, {Brogi}, {Malavolta}, {Wyttenbach}, {Line}, {Hoeijmakers}, {Fossati}, {Bonomo}, {Nascimbeni}, {Panwar}, {Affer}, {Benatti}, {Biazzo}, {Bignamini}, {Borsa}, {Carleo}, {Claudi}, {Cosentino}, {Covino}, {Damasso}, {Desidera}, {Giacobbe}, {Harutyunyan}, {Lanza}, {Leto}, {Maggio}, {Maldonado}, {Mancini}, {Micela}, {Molinari}, {Pagano}, {Piotto}, {Poretti}, {Rainer}, {Scandariato}, {Sozzetti}, {Allart}, {Borsato}, {Bruno}, {Di Fabrizio}, {Ehrenreich}, {Fiorenzano}, {Frustagli}, {Lavie}, {Lovis}, {Magazz{\`u}}, {Nardiello}, {Pedani}, \& {Smareglia}}]{Pino20}
{Pino}, L., {D{\'e}sert}, J.-M., {Brogi}, M., {et~al.} 2020, \apjl, 894, L27, \dodoi{10.3847/2041-8213/ab8c44}

\bibitem[{{Prinoth} {et~al.}(2022){Prinoth}, {Hoeijmakers}, {Kitzmann}, {Sandvik}, {Seidel}, {Lendl}, {Borsato}, {Thorsbro}, {Anderson}, {Barrado}, {Kravchenko}, {Allart}, {Bourrier}, {Cegla}, {Ehrenreich}, {Fisher}, {Lovis}, {Guzm{\'a}n-Mesa}, {Grimm}, {Hooton}, {Morris}, {Oreshenko}, {Pino}, \& {Heng}}]{Prinoth22}
{Prinoth}, B., {Hoeijmakers}, H.~J., {Kitzmann}, D., {et~al.} 2022, Nature Astronomy, 6, 449, \dodoi{10.1038/s41550-021-01581-z}

\bibitem[{{Prinoth} {et~al.}(2023){Prinoth}, {Hoeijmakers}, {Pelletier}, {Kitzmann}, {Morris}, {Seifahrt}, {Kasper}, {Korhonen}, {Burheim}, {Bean}, {Benneke}, {Borsato}, {Brady}, {Grimm}, {Luque}, {St{\"u}rmer}, \& {Thorsbro}}]{Prinoth23}
{Prinoth}, B., {Hoeijmakers}, H.~J., {Pelletier}, S., {et~al.} 2023, arXiv e-prints, arXiv:2308.04523, \dodoi{10.48550/arXiv.2308.04523}

\bibitem[{{Prinoth} {et~al.}(2024){Prinoth}, {Hoeijmakers}, {Morris}, {Lam}, {Kitzmann}, {Sedaghati}, {Seidel}, {Lee}, {Thorsbro}, {Borsato}, {Damasceno}, {Pelletier}, \& {Seifahrt}}]{Prinoth24}
{Prinoth}, B., {Hoeijmakers}, H.~J., {Morris}, B.~M., {et~al.} 2024, arXiv e-prints, arXiv:2403.08863, \dodoi{10.48550/arXiv.2403.08863}

\bibitem[{{Smith} {et~al.}(2024){Smith}, {Line}, {Bean}, {Brogi}, {August}, {Welbanks}, {Desert}, {Lunine}, {Sanchez}, {Mansfield}, {Pino}, {Rauscher}, {Kempton}, {Zalesky}, \& {Fowler}}]{Smith24}
{Smith}, P. C.~B., {Line}, M.~R., {Bean}, J.~L., {et~al.} 2024, \aj, 167, 110, \dodoi{10.3847/1538-3881/ad17bf}

\bibitem[{{Snellen} {et~al.}(2010){Snellen}, {de Kok}, {de Mooij}, \& {Albrecht}}]{Snellen10}
{Snellen}, I. A.~G., {de Kok}, R.~J., {de Mooij}, E. J.~W., \& {Albrecht}, S. 2010, \nat, 465, 1049, \dodoi{10.1038/nature09111}

\bibitem[{{Spring} {et~al.}(2022){Spring}, {Birkby}, {Pino}, {Alonso}, {Hoyer}, {Young}, {Coelho}, {Nespral}, \& {L{\'o}pez-Morales}}]{Spring22}
{Spring}, E.~F., {Birkby}, J.~L., {Pino}, L., {et~al.} 2022, \aap, 659, A121, \dodoi{10.1051/0004-6361/202142314}

\bibitem[{{Sreejith} {et~al.}(2023){Sreejith}, {France}, {Fossati}, {Koskinen}, {Egan}, {Cauley}, {Cubillos}, {Ambily}, {Huang}, {Panayotis Lavvas}, {Fleming}, {Desert}, {Nell}, {Petit}, \& {Vidotto}}]{Sreejith23}
{Sreejith}, A.~G., {France}, K., {Fossati}, L., {et~al.} 2023, arXiv e-prints, arXiv:2308.05726, \dodoi{10.48550/arXiv.2308.05726}

\bibitem[{{Stangret} {et~al.}(2022){Stangret}, {Casasayas-Barris}, {Pall{\'e}}, {Orell-Miquel}, {Morello}, {Luque}, {Nowak}, \& {Yan}}]{Stangret22}
{Stangret}, M., {Casasayas-Barris}, N., {Pall{\'e}}, E., {et~al.} 2022, \aap, 662, A101, \dodoi{10.1051/0004-6361/202141799}

\bibitem[{{Stock} {et~al.}(2018){Stock}, {Kitzmann}, {Patzer}, \& {Sedlmayr}}]{Stock18}
{Stock}, J.~W., {Kitzmann}, D., {Patzer}, A. B.~C., \& {Sedlmayr}, E. 2018, \mnras, 479, 865, \dodoi{10.1093/mnras/sty1531}

\bibitem[{{Tamuz} {et~al.}(2005){Tamuz}, {Mazeh}, \& {Zucker}}]{Tamuz05}
{Tamuz}, O., {Mazeh}, T., \& {Zucker}, S. 2005, \mnras, 356, 1466, \dodoi{10.1111/j.1365-2966.2004.08585.x}

\bibitem[{{Tan} \& {Komacek}(2019)}]{Tan19}
{Tan}, X., \& {Komacek}, T.~D. 2019, \apj, 886, 26, \dodoi{10.3847/1538-4357/ab4a76}

\bibitem[{Virtanen {et~al.}(2020)Virtanen, Gommers, Oliphant, Haberland, Reddy, Cournapeau, Burovski, Peterson, Weckesser, Bright, {van der Walt}, Brett, Wilson, Millman, Mayorov, Nelson, Jones, Kern, Larson, Carey, Polat, Feng, Moore, {VanderPlas}, Laxalde, Perktold, Cimrman, Henriksen, Quintero, Harris, Archibald, Ribeiro, Pedregosa, {van Mulbregt}, \& {SciPy 1.0 Contributors}}]{scipy}
Virtanen, P., Gommers, R., Oliphant, T.~E., {et~al.} 2020, Nature Methods, 17, 261, \dodoi{10.1038/s41592-019-0686-2}

\bibitem[{{Yan} {et~al.}(2020){Yan}, {Pall{\'e}}, {Reiners}, {Molaverdikhani}, {Casasayas-Barris}, {Nortmann}, {Chen}, {Molli{\`e}re}, \& {Stangret}}]{Yan20}
{Yan}, F., {Pall{\'e}}, E., {Reiners}, A., {et~al.} 2020, \aap, 640, L5, \dodoi{10.1051/0004-6361/202038294}

\bibitem[{{Yan} {et~al.}(2022){Yan}, {Pall{\'e}}, {Reiners}, {Casasayas-Barris}, {Cont}, {Stangret}, {Nortmann}, {Molli{\`e}re}, {Henning}, {Chen}, \& {Molaverdikhani}}]{Yan22}
---. 2022, \aap, 661, L6, \dodoi{10.1051/0004-6361/202243503}

\end{thebibliography}
\bibliographystyle{aasjournal}

\end{document}